\documentclass[pre,aps,nofootinbib,eqsecnum]{revtex4}
\usepackage{setspace}

\usepackage{amsmath,amssymb,bm}
\usepackage[colorlinks]{hyperref}
\usepackage[all]{hypcap}
\usepackage{graphicx,psfrag,float,epsfig}
\usepackage{mathrsfs,wasysym}
\usepackage{epstopdf}
\usepackage{comment}
\usepackage{pbox}
\usepackage{array}
\usepackage{xspace}
\usepackage[usenames,dvipsnames]{color}
\usepackage{sidecap,wrapfig}
\usepackage[applemac]{inputenc}

\DeclareMathAlphabet\mathbfcal{OMS}{cmsy}{b}{n}

\DeclareSymbolFontAlphabet{\mathrsfs}{rsfs}
\DeclareMathAlphabet{\mathcal}{OMS}{cmsy}{m}{n}

\DeclareSymbolFont{bbold}{U}{bbold}{m}{n}
\DeclareSymbolFontAlphabet{\mathbbold}{bbold}

\newcommand{\ud}{\mathrm{d}}
\allowdisplaybreaks[1]

\begin{document}

\title{Analytic Solutions to Compact Binary Inspirals With Leading Order Spin-Orbit Contribution Using The Dynamical Renormalization Group}


\def\Pitt{Pittsburgh Particle Physics Astrophysics and Cosmology Center (PITT PACC) \\ Department of Physics and Astronomy, University of Pittsburgh, Pittsburgh, Pennsylvania 15260, USA}

\author{Zixin Yang}
\email[E-mail:]{ziy8@pitt.edu}
\affiliation{\Pitt}
\author{Adam K. Leibovich}
\email[E-mail:]{akl2@pitt.edu}
\affiliation{\Pitt}

\begin{abstract}
	We calculate the real-space trajectory and spin precession of a generic spinning compact binary inspiral at any time instant using the dynamical renormalization group formalism. This method leads to closed-form analytic solutions to the binary motion through treating radiation reaction as perturbations and resumming the secular growth of perturbative terms. We consider the spin-orbit effects at leading order and the 2.5PN radiation reaction without orbit averaging or precession averaging for arbitrary individual masses and spin magnitudes and orientations. The solutions are written in a moving reference frame, with the orbital angular momentum and binary radial directions aligned along two of the axes. The resummed solutions show improved accuracy compared to adiabatic solutions while also being an order of magnitude faster computationally compared to numerical integration methods.
\end{abstract}

\maketitle

\section{Introduction}

A worldwide network of gravitational wave (GW) detectors is being developed to monitor the ripples in the fabric of spacetime passing through the earth. This includes the ground-based laser interferometers GEO600, LIGO and VIRGO collaborations currently in operation, and the under-construction spaced-based observatory LISA and cryogenic detector KAGRA in Japan.  The successful detection of gravitational waves from inspiraling black holes (BH) and neutron stars (NS) by the LIGO and VIRGO collaborations directly and spectacularly confirmed one of the predictions of Einstein's Theory of General Relativity. A generic prediction of metric theories of gravity, BH or NS coalescence is a strong source of GWs for interferometric detectors. To successfully identify and analyze the gravitational wave signals, it is necessary to construct a systematic description of the binary black hole dynamics and waveforms during coalescence. A set of expected waveforms portrayed by the intrinsic parameters of the compact binary within the astrophysically interesting region of the parameter space forms a waveform template bank \cite{Babak:2006ty, Ajith:2007kx, Harry:2016ijz}. Using these precise waveform templates, a matched filtering technique is used to try to discover the weak GW signals buried in the detector noise. More accurate templates will help us extract physical information from the observed events in order to gain further knowledge of the black hole or neutron star properties. 


The choice of the BH coalescence model is critical for determining the waveform. The last few orbits of the inspiral phase through the merger and ringdown of the BH coalescence have been simulated by Numerical Relativity \cite{Baker:2006ha, Centrella:2010mx}. There have been developments on the analytic understandings for merger and ringdown \cite{, McWilliams:2018ztb}. The slowly-orbiting long inspiral phase can be studied analytically using post-Newtonian (PN) perturbation theory with small velocity and weak field approximations. BH dynamics is described by the Newtonian-like equations of motion in the form of the acceleration of the binary constituents.  During the inspiral, the binary slowly loses energy and angular momentum to gravitational radiation starting at 2.5PN \cite{Burke:1970dnm,thorne1985laws}. Higher order corrections up to 4PN in the conservative sector have been calculated \cite{Damour:2014jta,Damour:2016abl,Jaranowski:2015lha,Bernard:2015njp,Bernard:2017bvn,Marchand:2017pir,Bernard:2017ktp,Foffa:2019rdf,Foffa:2019yfl}. Solving for the motions is the fundamental step in obtaining the waveforms and deriving the evolution of the theoretical physical measurements in time, such as the GW phase directly measured by the detectors and power loss due to gravitational radiations. 

The exact solutions to the motions can be found by numerically integrating these nonlinearly-coupled ordinary differential equations. However, in calculations of template banks, each point in the intrinsic parameter space representing a waveform with different initial conditions requires a new numerical computation. The sample rate of the corresponding waveform directly depends on the precision and step sizes of the solutions of the motions. The discrete nature of the computational solutions also brings the issue of the distance between the templates in the parameter space, which may result in the loss in signal-to-noise ratio due to the mismatch of the template in the match-filter of the signal data. Since the third observing run of LIGO and Virgo are having a weekly rate of observed events, a faster and more accurate way of computation in the signal analysis is critical, with even larger rates expected with future upgrades. A  fully analytic waveform solution with continuous parameters would certainly increase calculation efficiency for template-based data analysis.

The adiabatic approximation is often used to find the analytic solutions to the motion, including  inspiral radiation reaction effects \cite{Drasco:2005is, Hughes:2005qb, Sago:2005gd}. Using the PN expansions of the conserved energy $E$ and flux $\mathcal{F}$, the adiabatic waveforms are obtained by solving the energy balance equation $\ud E / \ud t = \mathcal{F}$. The balance equation leads to the secular evolutions of the orbital angular frequency $\omega(t)$, from which one can derive the accumulated phase of gravitational waves $\phi(t)=2\int \ud \tau \omega(\tau) $. An implicit assumption in the energy balance equation is that $E$ does not change much over an orbital timescale. In other words, the adiabatic solutions are orbit-averaged and thus remove some of the orbital detail. The adiabatic approximation fails to account for  secular evolution of some of the orbital elements, which can lead to measurable phasing effects \cite{Pound:2005fs}.

When considering spinning black holes, which adds 6 additional degrees of freedom, the binary motions become more complicated. The convention of the PN order counting of the spin here is defined as $|\bm{S}| = \chi m^2$, where $m$ is the mass of the object and $\chi$ a dimensionless spin parameter. For a maximally rotating compact BHs, $\chi \sim 1$. The leading contributions from spin-orbit effects enter into the motion at 1.5PN and spin-spin at 2PN, before the leading order radiation reaction force. The major effect of the presence of the spin on the orbital evolution is that a spin component perpendicular to the orbital angular momentum causes the orbital plane to precess. This means the orbital plane will change its orientation when it is not perpendicular to the spin vector. Thus the observed waveform, depending on the orbital orientation with respect to the detector, will modulate due to spin-induced orbital precession. The secular evolutions of the spins themselves are given by the spin precession equations \cite{barker1979gravitational, Thorne:1984mz}. With the spin precession equations, it is possible to determine the angular momentum transfer between orbital and spin angular momenta and the total angular momentum loss during the inspiral regime. One of the recent works to construct analytic spin-precessing inspirals is through multiple scale analysis \cite{Chatziioannou:2016ezg, Chatziioannou:2017tdw}. This method gives orbit-averaged and precession-averaged closed form solutions by making a clean separation among the orbital time, precession time, and radiation reaction time scales and treating the physical parameters by averaging over the longer time scales to solve for the shorter ones.  However, any averaging procedure results in the loss of some of the orbital dynamics.

In order to find analytic solutions to the spinning binary equations of motion and spin precession equations without any averaging  procedures, we follow the Dynamical Renormalization Group (DRG) formalism proposed by Galley and Rothstein in \cite{Galley:2016zee}. The idea of the DRG method is based on  renormalization group theory and the resummation of the singularities for perturbative ordinary differential equation problems \cite{Chen:1995ena}. The DRG method applied to binary inspirals starts by treating some of the higher PN order radiation reaction terms as a perturbation to a conservative background orbit. The secular growths of the perturbations are then resummed to preserve the correct power counting of the perturbations. In their work, Galley and Rothstein calculated the resummed solution for a non-spinning binary with leading order radiation up to the second-order corrections and included the PN corrections to the radiation reaction force. In this paper we incorporate spin-orbit effects and the leading-order radiation reaction, using the DRG method to obtain real-time solutions to the generic precessing compact binaries. 

The organization of this paper is as follows: In Section \ref{section_setup} we introduce the PN equations of motion and spin precession equations for compact binary inspirals. We also set up a moving coordinate frame using the radial vector and orbital angular momentum vector in which we will present our solutions. In Section \ref{section_DRG_solution} we summarize the procedures of the DRG method and give the resulting closed-form analytic solutions to the binary motions and spin precession. In Section \ref{section_num_compare}, we compare our DRG resummed solutions to the numerical and adiabatic solutions of the same equations. We also show a rough comparison of the calculation run time between the numerical integration and resummed solution substitution. We conclude in Section \ref{section_Conclusion}.
In Appendix \ref{appendix_orb_solutions} and \ref{appendix_prec_solutions}, we present the detailed calculations of the DRG method for orbital motions and spin precession, respectively. In Appendix \ref{appendix_frame} we propose a naive transformation of the moving coordinate frame to a fixed observer frame for the purpose of waveform construction.

\section{Leading Order Spin-Orbit Equations of Motion And Spin Precession Equations}
\label{section_setup}

The equations of motion of the compact binaries in the center-of-mass frame, including the Newtonian order, the leading-order spin-orbit contributions at 1.5PN in covariant spin supplementary condition (SSC), and the Burke-Thorne term due to the radiation-reaction force at 2.5PN, are given by \cite{Kidder:1995zr, Burke:1970dnm, Blanchet:2013haa}
\begin{align}
	\bm{a} = \bm{a}_{\textrm{N}} + \bm{a}_{\textrm{SO}} + \bm{a}_{\textrm{RR}}, \label{a_sum}
\end{align}
where the terms in the post-Newtonian hierarchy are
\begin{subequations}
\begin{align}
		\bm{a}_{\textrm{N}} &= - \frac{M}{r^2} \hat{\bm{n}}, \\
		\bm{a}_{\textrm{SO}} &= \frac{1}{r^3}\Bigg\{6\hat{\bm{n}} \Bigg[\big(\hat{\bm{n}} \times \bm{v}\big) \cdot \Bigg(2\bm{S} +\Delta \bm{\Sigma}  \Bigg)   \Bigg] - \Bigg[ \bm{v} \times \Bigg(7\bm{S} + 3 \Delta \bm{\Sigma}  \Bigg) \Bigg] + 3\dot{r} \Bigg[\hat{\bm{n}} \times \Bigg(3\bm{S} + \Delta \bm{\Sigma}  \Bigg) \Bigg] \Bigg\}, \label{a_so}\\
	\bm{a}_{\textrm{RR}} &= \frac{M^2 \nu}{15 r^4}\dot{r}\Big(\frac{136M}{r} +72 \bm{v}^2\Big)\bm{r} - \frac{8M^2\nu}{5r^3}\Big(\frac{3M}{r} + \bm{v}^2 \Big)\bm{v}. 
\end{align}	
\end{subequations}
In the expressions above, $\bm{r}$ and $\bm{v}$ are the binary relative center-of-mass separation and velocity, $\hat{\bm{n}} \equiv \bm{r}/r$ and $\dot{r} = \ud r / \ud t = \hat{\bm{n}}\cdot\bm{v} $. The binary masses are denoted as $m_{1,2}$, the total binary mass $M=m_1 + m_2$, $\nu \equiv m_1 m_2/M^2$ and $\Delta \equiv (m_1 - m_2)/M$. The combinations of the individual spins are written as
	\begin{align}
		\bm{S} = \bm{S}_1 +\bm{S}_2, \qquad \bm{\Sigma} = \frac{\bm{S}_2}{X_2} - \frac{\bm{S}_1}{X_1} ,
	\end{align}
with $X_a = m_a/M$. The spin vectors precess due to spin-orbit coupling following the relation of \cite{thorne1985laws, Kidder:1995zr}
	\begin{align}
		\dot{\bm{S}}_a &= \frac{1}{r^3}\big(\bm{L}_N \times \bm{S}_a\big)\Bigg(2 + \frac{3}{2}\frac{m_b}{m_a} \Bigg)  \label{spinprec},
	\end{align}
where $\{a,b\}$ are the binary labels $\{1,2\}$, and $\bm{L}_N = \nu M (\bm{r} \times \bm{v})$ is the Newtonian orbital angular momentum. \\

In order to obtain the analytic solutions to the inspiral equations of motions and the spin precession equations (\ref{a_sum})-(\ref{spinprec}), we adopt a coordinate frame $\{\bm{\bm{n}, \bm{\lambda}, \bm{l}}\}$,
moving along with the center-of-mass and the orientation of its motion \cite{Blanchet:2013haa, arun2009higher, marsat2013next}, where $\bm{l} = \bm{n} \times \bm{v}/|\bm{n} \times \bm{v}|$ and $\bm{\lambda} = \bm{l} \times \bm{n}$ to complete an orthonormal basis triad. In this moving basis, the relative velocity can be expressed as
\begin{align}
	\bm{v} = \dot{r} \bm{n} + r\omega\bm{\lambda}
\end{align}
where $\omega$ is the orbital angular frequency of the binary. The relative acceleration $\bm{a} = \ud \bm{v}/\ud t$ in the moving basis is
\begin{align}
	\bm{a} = (\ddot{r} - r\omega^2)\bm{n} + (r\dot{\omega}+2\dot{r}\omega)\bm{\lambda}+r\varpi\omega\bm{l} \label{a_decomp},
\end{align}
where the orbital plane precession $\varpi$ of the orbit is defined as $\varpi \equiv -\bm{\lambda}\cdot\ud \bm{l}/\ud t$.

In terms of the moving basis components, the equations of motions (\ref{a_sum}) are
\begin{subequations}
\begin{align}
	\ddot{r} -r\omega^2 =& -\frac{M}{r^2} + \frac{64M^3\nu}{15r^4}\dot{r} + \frac{16M^2\nu}{5r^3}\dot{r}^3 + \frac{16M^2\nu}{5r}\dot{r}\omega^2 + \frac{\omega}{r^2}\big(5S_l + 3\Delta\Sigma_l\big),\label{EoM_r}\\
	r\dot{\omega}+2\dot{r}\omega = &-\frac{24M^3\nu}{5r^3}\omega - \frac{8M^2\nu}{5r^2}\dot{r}^2\omega -\frac{8M^2\nu}{5}\omega^3 -\frac{2\dot{r}}{r^3}S_l, \label{EoM_omega}\\
	\varpi =& \frac{2\dot{r}}{r^4\omega}S_{\lambda} + \frac{7}{r^3}S_n + \frac{3\Delta}{r^3}\Sigma_n \label{varpi},
\end{align}
\end{subequations}
where we decompose the spin $\bm{S} = S_n \bm{n} + S_{\lambda} \bm{\lambda} +S_l \bm{l}$, and similarly for $\bm{\Sigma}$. The spin precession equations (\ref{spinprec}) become
\begin{subequations}
\begin{align}
	\frac{\ud S^{a}_{n}}{\ud t} &= \Big(\omega - \Omega_a  \Big) S^{a}_{\lambda},  \label{S_n_prec}\\
	\frac{\ud S^{a}_{\lambda}}{\ud t} &= -\Big(\omega - \Omega_a  \Big) S^{a}_{n} + \varpi S^{a}_{l}, \\
	\frac{\ud S^{a}_{l}}{\ud t} &= -\varpi S^{a}_{\lambda}, \label{S_l_prec}
\end{align}
\end{subequations}
where we denote
\begin{align}
 	\Omega_{a} \equiv \frac{\nu M\omega}{r}\Bigg(2 + \frac{3}{2}\frac{m_b}{m_a}\Bigg) \label{Omega_a},
 \end{align} 
which is the norm of the precession vector of the $a$-th spin. The precession frequency $\varpi$, explicitly given by (\ref{varpi}), is of order $\mathcal{O}(S)$. At linear order in spin, the precession equations become
\begin{subequations}
	\label{spin_prec_decomp}
\begin{align}
	\frac{\ud S^{a}_{n}}{\ud t} &= \Big(\omega - \Omega_a  \Big) S^{a}_{\lambda},\\
	\frac{\ud S^{a}_{\lambda}}{\ud t} &= -\Big(\omega - \Omega_a  \Big) S^{a}_{n} + \mathcal{O}(S^2), \\
	\frac{\ud S^{a}_{l}}{\ud t} &= \mathcal{O}(S^2). 
\end{align}
\end{subequations}
Thus at  order  $\mathcal{O}(S)$, the $l$-component of the spin vectors are invariant, which are also the only components that appears in the orbital equations of motion in (\ref{EoM_r}) and (\ref{EoM_omega}). In solving these two equations by the DRG method, we then are able to treat $S_l$ and $\Sigma_l$ as time-independent constants. Following Ref~\cite{Galley:2016zee}, here we ignore the 1PN and 2PN conservative forces, as well as the next-to-leading order spin-orbit effects, which is the same order in the Post-Newtonian expansion as the 2.5PN radiation reaction terms. Instead, we focus on the leading order radiation reaction effects on spinning objects. In order to obtain gravitational wave templates, to be consistent we would need to include at least the 1PN conservative forces.    \\

\section{DRG Solutions to Dynamics and Spin Precession}
\label{section_DRG_solution}

The background quasi-circular orbit of a conserved binary with Newtonian and leading spin-orbit effects can be described by
\begin{equation}
     \Omega^2_B = \frac{M}{R^3_B} - \frac{\Omega_B}{R_B^3}\big(5S_l + 3\Delta\Sigma_l\big),
 \end{equation} 
with constant radius $R_B$ and constant angular frequency $\Omega_B$. To include the radiation reaction as perturbative effects, we write the orbital solutions as 
\begin{equation}
    r(t) = R_B + \delta r(t) + \delta r_S (t), \qquad \omega(t) = \Omega_B + \delta \omega(t) + \delta \omega_S (t), 
\end{equation}
where the first time-dependent terms $\delta r(t)$ and $\delta \omega(t)$ are the perturbation due to the 2.5PN radiation reaction force without the spin at a given time $t$. The $\delta r_S(t)$ and $\delta \omega_S(t)$ represent the perturbations due to the interaction of 1.5PN spin effects and the 2.5PN radiation reaction. The power counting at the initial time $t_0$ for each perturbation is given by
\begin{align}
    \delta r \sim v^5 R_B, \qquad \delta \omega \sim  v^6/R_B, \qquad \delta r_S \sim S v^4 /R_B, \qquad \delta \omega_S \sim S v^5/R_B^3
\end{align}
where we keep the spin as a placeholder expansion parameter instead of converting to PN orders for generality. Substituting the perturbed orbital radius and frequency into the equations of motion (\ref{EoM_r}) and (\ref{EoM_omega}), we find the solutions to the perturbation
\begin{subequations}
\label{del_sol}
\begin{align}
    \delta r_S(t) =& -\Big(\frac{144}{5}S_l + 48\Delta\Sigma_l \Big)\nu R_B^3 \Omega_B^5 (t-t_0) + \frac{\big(7S_l +3\Delta\Sigma_l\big)}{2\Omega_B R_B^3}A_B \Big[2\Omega_B(t-t_0)\cos{(\Omega_B(t-t_0)+\Phi_B)}\nonumber\\
    & \qquad\qquad\qquad - \sin{(\Omega_B(t-t_0)+\Phi_B)} \Big] + A^S_B \cos{(\Omega_B(t-t_0)+\Phi_B)}, \\
    \delta\omega_S(t) = &\Big(-\frac{24}{5}S_l + \frac{216}{5}\Delta \Sigma_l \Big)\nu R^2_B \Omega_B^6 (t-t_0) + \big(5S_l + 3\Delta \Sigma_l \big)\frac{A_B}{R_B^4}\sin{(\Omega_B(t-t_0)+\Phi_B)} \nonumber \\
    &- \big(14S_l + 6\Delta \Sigma_l \big)\frac{A_B}{R_B^4}\Omega_B(t-t_0)\cos{(\Omega_B(t-t_0)+\Phi_B)} - \frac{2A^S_B \Omega_B}{R_B} \cos{(\Omega_B(t-t_0)+\Phi_B)}. 
\end{align}
and also the time integration of $\delta\omega_S(t)$, $\delta\Phi_S(t)$, which is the perturbation of the orbital phase $\phi(t)$,
\begin{align}
    \delta\Phi_S(t) = &\Big(-\frac{12}{5}S_l + \frac{108}{5}\Delta \Sigma_l \Big)\nu R^2_B \Omega_B^6 (t-t_0)^2 - \Big(19S_l + 9\Delta \Sigma_l \Big)\frac{A_B}{\Omega_B R_B^4}\cos{(\Omega_B(t-t_0)+\Phi_B)} \nonumber \\
    & - \Big(14S_l + 6\Delta \Sigma_l \Big)\frac{A_B}{R_B^4}(t-t_0)\sin{(\Omega_B(t-t_0)+\Phi_B)} - \frac{2A^S_B}{R_B} \sin{(\Omega_B(t-t_0)+\Phi_B)},
\end{align}
\end{subequations}
where $\Phi_B$, $A_B$, and $A_B^S$ are integration constants. $\{R_B, \Omega_B, \Phi_B, A_B, A_B^S \}$ forms a set of bare parameters to be determined by initial conditions. While $e_B= A_B / R_B$ is the small orbital eccentricity of order $\mathcal{O}(v^5)$ induced by the radiation reaction force, the interaction between the spin and radiation reaction lead to a smaller eccentricity $e^S_B = A^S_B/R_B \sim \mathcal{O}(Sv^4)$. The spin-radiation eccentricity deforms the circular orbit out-of-phase compared to the radiation eccentricity, although with a fixed phase difference. 

To maintain the power countings of the perturbations, the secularly growing terms in (\ref{del_sol}) are absorbed into the bare parameters through the relations
\begin{subequations}
\begin{align}
    R_B (t_0) &= R_R(\tau) + \delta_R^{v^5}(\tau,t_0) + \delta_R^{S}(\tau,t_0),\\
    \Omega_B (t_0) &= \Omega_R(\tau) + \delta_\Omega^{v^5}(\tau,t_0) + \delta_\Omega^{S}(\tau,t_0),\\
    \Phi_B (t_0) &= \Phi_R(\tau) + \delta_\Phi^{v^5}(\tau,t_0) + \delta_\Phi^{S}(\tau,t_0),\\
    A^S_B (t_0) &= A^S_R(\tau) + \delta_A^{S}(\tau,t_0),
\end{align}
\end{subequations}
where $\{R_R, \Omega_R, \Phi_R, A_R^S\}$ are  the ``renormalized" parameters depending on an arbitrary renormalization scale $\tau$. The quantities $\{\delta_R^{v^5}, \delta_R^{S}, \delta_\Omega^{v^5}, \delta_\Omega^{S}, ...\}$ are  counter-terms, to be determined by renormalizing the perturbation expansions. Introducing the renormalization scale into the perturbation solutions (\ref{del_sol}) by writing $t-t_0 = (t - \tau) + (\tau -t_0)$ and using the counter-terms to cancel all the secular $(\tau - t_0)$ terms, we find that   
\begin{subequations}
\label{EOM_resummed}
\begin{align}
    r(t) =& R_R(t) + \Bigg(1 - \frac{\big(7S_l +3\Delta\Sigma_l\big)}{2\Omega_R(t) R_R^3(t)}\Bigg)A_R(t) \sin{\Phi_R(t)} + A^S_R(t)\cos{\Phi_R(t)},\\
    \omega(t) =& \Omega_R(t) - \frac{2\Omega_R(t)A_R(t)}{R_R(t)}\Bigg(1 - \frac{\big(5S_l +3\Delta\Sigma_l\big)}{2\Omega_R(t) R_R^3(t)}\Bigg)\sin{\Phi_R(t)} - \frac{2A^S_R \Omega_R(t)}{R_R(t)} \cos{\Phi_R(t)},\\
    \phi(t) =& \Phi_R(t) + \frac{2A_R(t)}{R_R(t)}\Bigg(1 - \frac{\big(19S_l +9\Delta\Sigma_l\big)}{2\Omega_R(t) R_R^3(t)}\Bigg)\cos{\Phi_R(t)} - \frac{2A^S_R(t) }{R_R(t)} \sin{\Phi_R(t)},
\end{align}
\end{subequations}
where $r(t)$ and $\omega(t)$ are the orbital radius and frequency defined in the previous section, and $\phi(t)$ is the time integral of $\omega(t)$ representing the orbital phase of the binary inspiral. The renormalized parameters are determined at arbitrary time via the renomalization group equations, determined using the fact that the corresponding bare parameters are independent of the choice of $\tau$. The ``beta-functions'' of the RG equations are determined by the counter-terms, leading to the first-order equations satisfied by the renormalized parameters. We give the RG solutions in the form of invariance in time as 
\begin{subequations}
\label{RGE_inv}
\begin{align}
     \frac{64\nu M^3}{5}t +\frac{1}{4}R_R(t)^4 + \frac{2\mathscr{S}}{5M^{1/2}}R_R(t)^{5/2} + \frac{\mathscr{S}^2}{M}R_R(t) +\frac{2\mathscr{S}^{8/3}}{\sqrt{3}M^{4/3}}\tan^{-1}\Bigg(\frac{1}{\sqrt{3}} + \frac{2M^{1/6}R_R(t)^{1/2}}{\sqrt{3}\mathscr{S}^{1/3}}\Bigg) &\nonumber\\
    +\frac{\mathscr{S}^{8/3}}{3M^{4/3}}\ln\Bigg(\frac{\Big(\mathscr{S}^{1/3}-M^{1/6}R_R(t)^{1/2}\Big)^2}{\mathscr{S}^{2/3} + \mathscr{S}^{1/3}M^{1/6}R_R(t)^{1/2} + M^{1/3}R_R(t)} \Bigg) =\textrm{constant}&,\\\nonumber\\
    \Omega_R^2(t) R_R^3(t) + \Omega_R(t)\big(5S_l + 3\Delta\Sigma_l \big) =M&,\\\nonumber\\
    \Phi_R(t) + \frac{1}{32M^{5/2}\nu} R_R^{5/2}(t) -\frac{5\Big(41S_l + 15\Delta \Sigma_l \Big)}{256\nu M^{2}\mathscr{S}^{2}} \Bigg(\frac{64\nu M^3}{5}t +\frac{1}{4}R_R^4(t) + \frac{2\mathscr{S}}{5M^{1/2}}R_R^{5/2}(t)\Bigg) =\textrm{constant}&,\\\nonumber\\
    A_R(t) = \textrm{constant}&,\\\nonumber\\
    A^S_R(t) -\frac{5A_R\Big(7S_l + 3\Delta \Sigma_l \Big)}{64\nu M^{2}\mathscr{S}^{2}}\Bigg(\frac{64\nu M^3}{5}t +\frac{1}{4}R_R^4(t) + \frac{2\mathscr{S}}{5M^{1/2}}R_R^{5/2}(t)  \Bigg) =\textrm{constant}&,
 \end{align} 
\end{subequations}
where for convenience we have defined $\mathscr{S} \equiv (51+21\Delta\Sigma_l)/4$. Remember, at this order the $l$-component of the spin vectors are constant. The constants in the equations above can be further determined using the initial conditions by solving (\ref{EOM_resummed}) at a given time instant. The expressions in (\ref{EOM_resummed}) combined with (\ref{RGE_inv}) give the resummed solution to the 0PN spinning inspiral dynamics valid up to times $(t-t_0)$ of order $1/(\nu v^5(t)\Omega_R(t))$. To improve the accuracy, we need to calculate higher order perturbations in the same formalism or include higher PN conservative corrections to the motions. 

The background solution to the conserved spin precessions has a constant precession frequency. We renormalize the precession frequency perturbed by the radiation reaction using the same DRG procedure. The resummed solutions to the spin precession equations (\ref{spinprec}) are 
\begin{align}
    S^{a}_{+}(t) = \mathcal{S}^{a}_{+R}(t) \exp \Bigg\{&i\Bigg[\frac{2A^S_R(t) }{R_R(t)} - \frac{3\nu_a}{\nu} A_R^S(t) \Omega_R(t)^2  R_R(t)\Bigg]\sin{\Phi_R(t)}\nonumber\\
    &-i\Bigg[\frac{2 A_R(t)}{R_R(t)} - \Big(19S_l + 9\Delta \Sigma_l \Big)\frac{A_R(t)}{R_R(t)^4\Omega_R(t)}- \frac{3\nu_a}{\nu} \Omega_R(t)^2 A_R(t) R_R(t)\nonumber\\
    &\qquad\quad+ \frac{\nu_a}{\nu}\Big(\frac{29}{2}S_l +\frac{9}{2}\Delta\Sigma_l\Big)\frac{A_R(t)\Omega_R(t) }{R_R(t)^2}\Bigg]\cos{\Phi_R (t)}\Bigg\} ,
\end{align}
where $ \nu_a\equiv (2+\frac{3m_b}{2m_a}) \nu^2  $ and $S^{a}_{+} \equiv S^{a}_{n} + i S^{a}_{\lambda}$ contains the two precessing components of the spin vector in the moving triad $\{\bm{n}, \bm{\lambda}, \bm{l}\}$ coordinate system. The exponential preserves the magnitudes of the spin vectors, which is  conserved as can easily be seen from Eq.~(\ref{spinprec}). The renormalized parameter $\mathcal{S}^{a}_{+R}(t)$ can be written in terms of invariance over time and other parameters as
\begin{align}
    i\ln \mathcal{S}^{a}_{+R}(t) - \Phi_R(t) -\frac{5\nu_a R^{3/2}_R(t)}{96M^{3/2}\nu^2} - \frac{5\Big(41S_l + 15\Delta \Sigma_l \Big)\nu_a}{384M^2\nu^2}\ln\big(M^{1/2}R_R^{3/2}(t) - \mathscr{S}\big)=\textrm{constant}.
\end{align}
We include the more detailed calculations and renormalization procedures in the appendices for interested readers.

\section{Numerical Solution Comparison}
\label{section_num_compare}

To compare our analytic solutions to the orbital equations of motion and the spin precession equations, we solve the sets of equations numerically and compare with the DRG solution solved with the same initial conditions. We choose to compare compact binary systems of total mass $M=1$. The initial conditions for the physical parameters are related to the renormalized parameters through the renormalized solutions (\ref{orbit_resummed}) and (\ref{spin_resummed}) at $t_i = 0$. We choose for our initial conditions
\begin{align}
    \label{orbital_data}
    \left. 
    \begin{array}{r c l}
        \Omega_R(0) & = & 10^{-2}/ M \\
        R_R (0) & = & \left( \displaystyle\frac{M}{\Omega_R(0)^2} - \displaystyle\frac{5S_l + 3\Delta\Sigma_l}{\Omega_R(0)} \right)^{1/3} \\
        \Phi_R(0) & = & 0 \\
        A^S_R(0) & = & 0 \\
        A_R(0) & = & \displaystyle\frac{\displaystyle\frac{64 \nu}{5} R_R(0)^6 \Omega_R(0)^5 + \displaystyle\frac{\nu }{5}   R_R(0)^3 \Omega_R (0)^4 (144 S_l + 240 \Delta \Sigma_l)}{ \left(1 + \displaystyle\frac{7 S_l + 3 \Delta \Sigma_l}{2 R_R(0)^3 \Omega_R (0)}\right)}
    \end{array}
    \right\}  
    ~~ \Longrightarrow ~~
    \left\{
    \begin{array}{r c l}
        r(0) & = & R_R(0) \\
        \dot{r}(0) & = & 0 \\
        \omega(0) & = & \Omega_R(0) \\
        \phi(0) & = & \displaystyle\frac{2 A_R(0)}{R_R(0)} \left(1-\frac{19 S_l + 9 \Delta \Sigma_l}{2 R_R(0)^3 \Omega_R(0)}\right)
    \end{array}
    \right.
\end{align} where the expression for $A_R(0)$ comes from
\begin{align}
    \dot{r}(t_i) =& A_R(t_i) \Omega_R (t_i) \cos \Phi_R(t_i) \left(1 + \frac{7 S_l + 3 \Delta \Sigma_l}{2 R_R(t_i)^3 \Omega_R (t_i)}\right) - \frac{64 \nu}{5} R_R(t_i)^6 \Omega_R (t_i)^6 \nonumber\\
    & - \frac{1}{5} \nu  R_R(t_i)^3 \Omega_R (t_i)^5 (144 S_l + 240 \Delta \Sigma_l) -A^S_R(t_i) \Omega_R (t_i) \sin \Phi_R(t_i) .
\end{align}
and $\dot{r}(0)$ is taken to be zero for quasi-circular motion. Meanwhile, we impose a small  non-vanishing $\mathcal{O}(v^5)$ eccentricity $e_R = A_R/R_R(t)$, and a spin-induced eccentricity $e^S_R = A^S_R(0)/R_R(0)$ at $\mathcal{O}(v^4S)$ that runs starting from zero.
\begin{figure}
    \includegraphics[width=0.48\columnwidth]{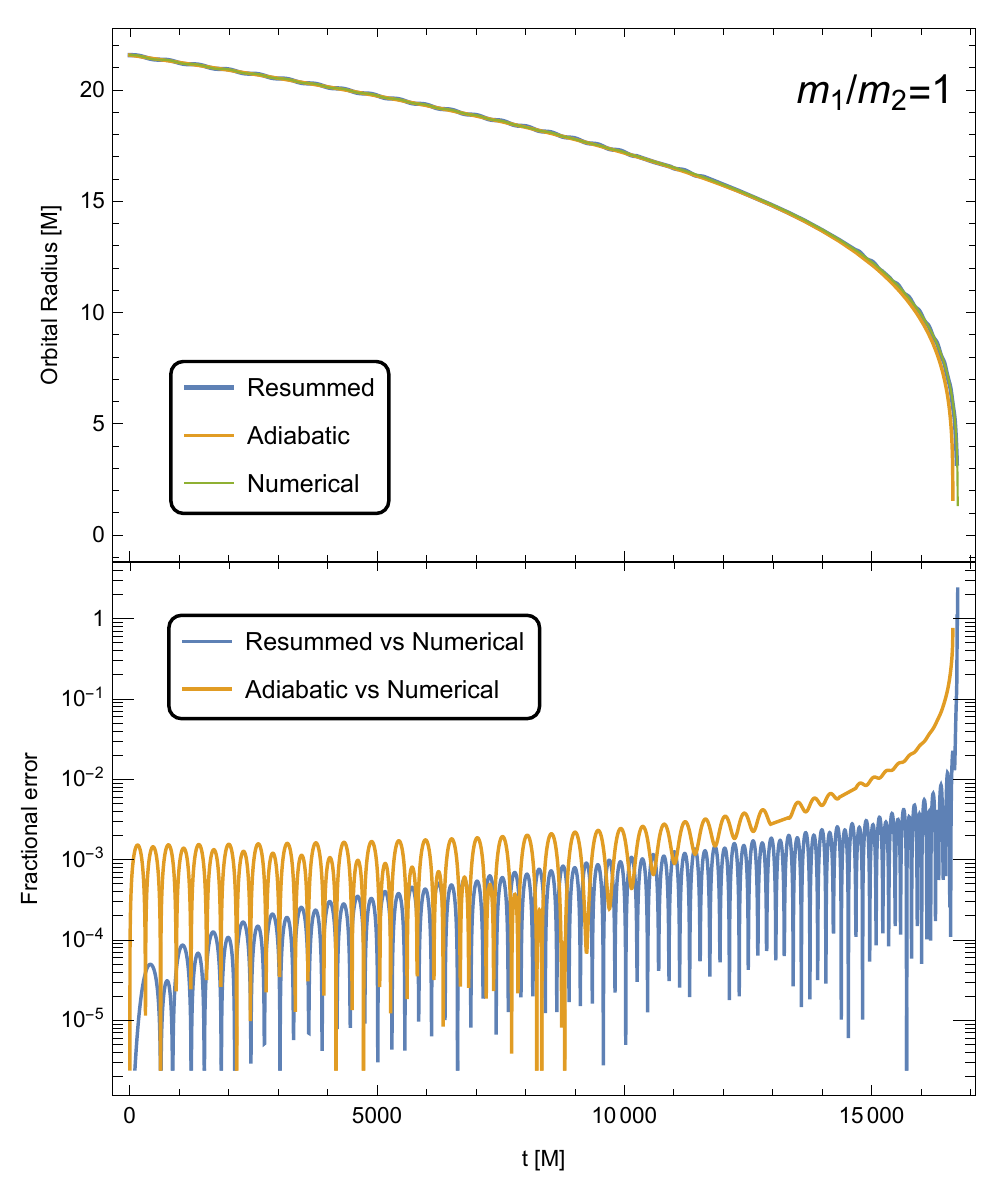}
    \includegraphics[width=0.48\columnwidth]{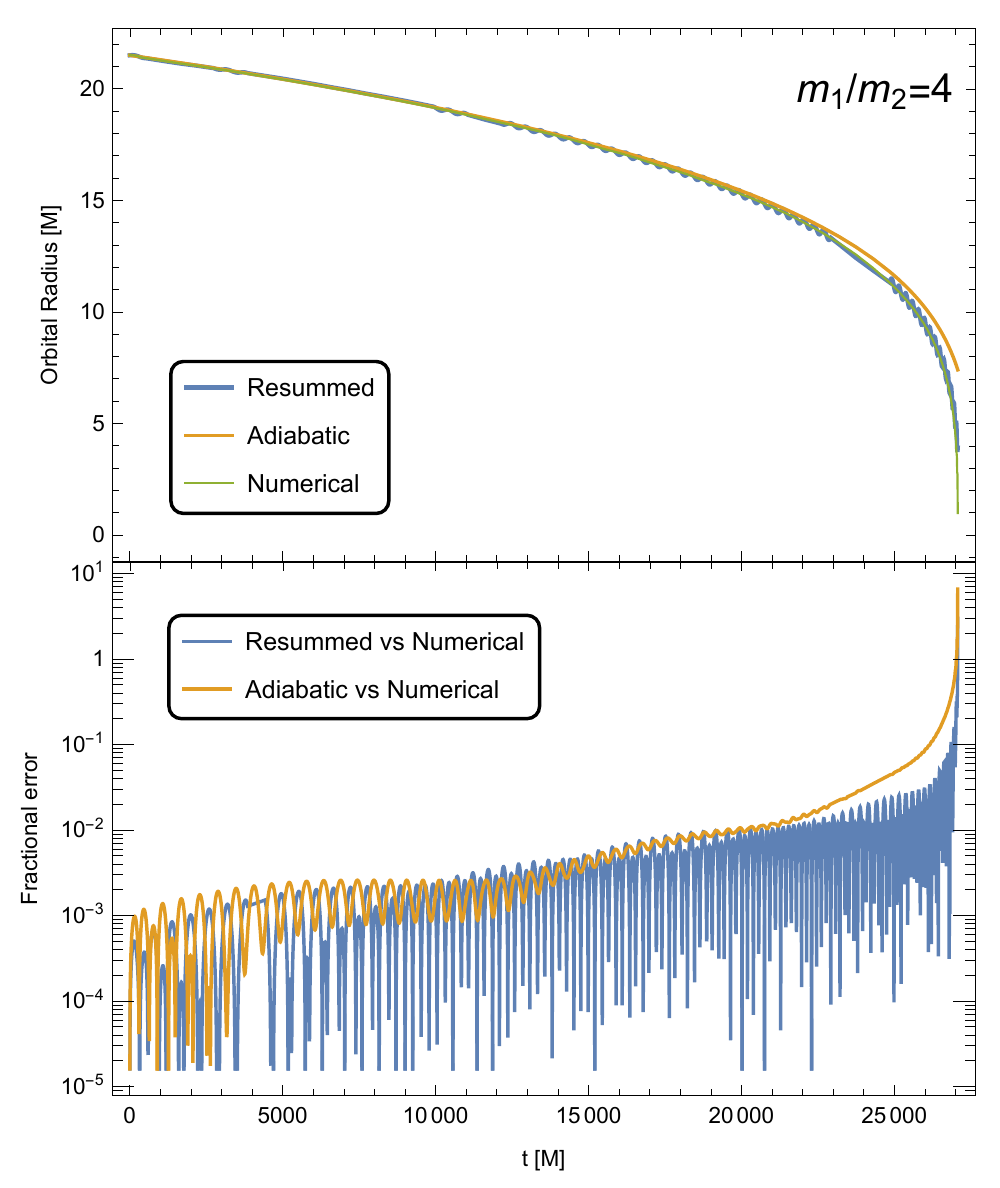}
    \includegraphics[width=0.48\columnwidth,trim = 0 0.1cm 0 0.2cm,clip]{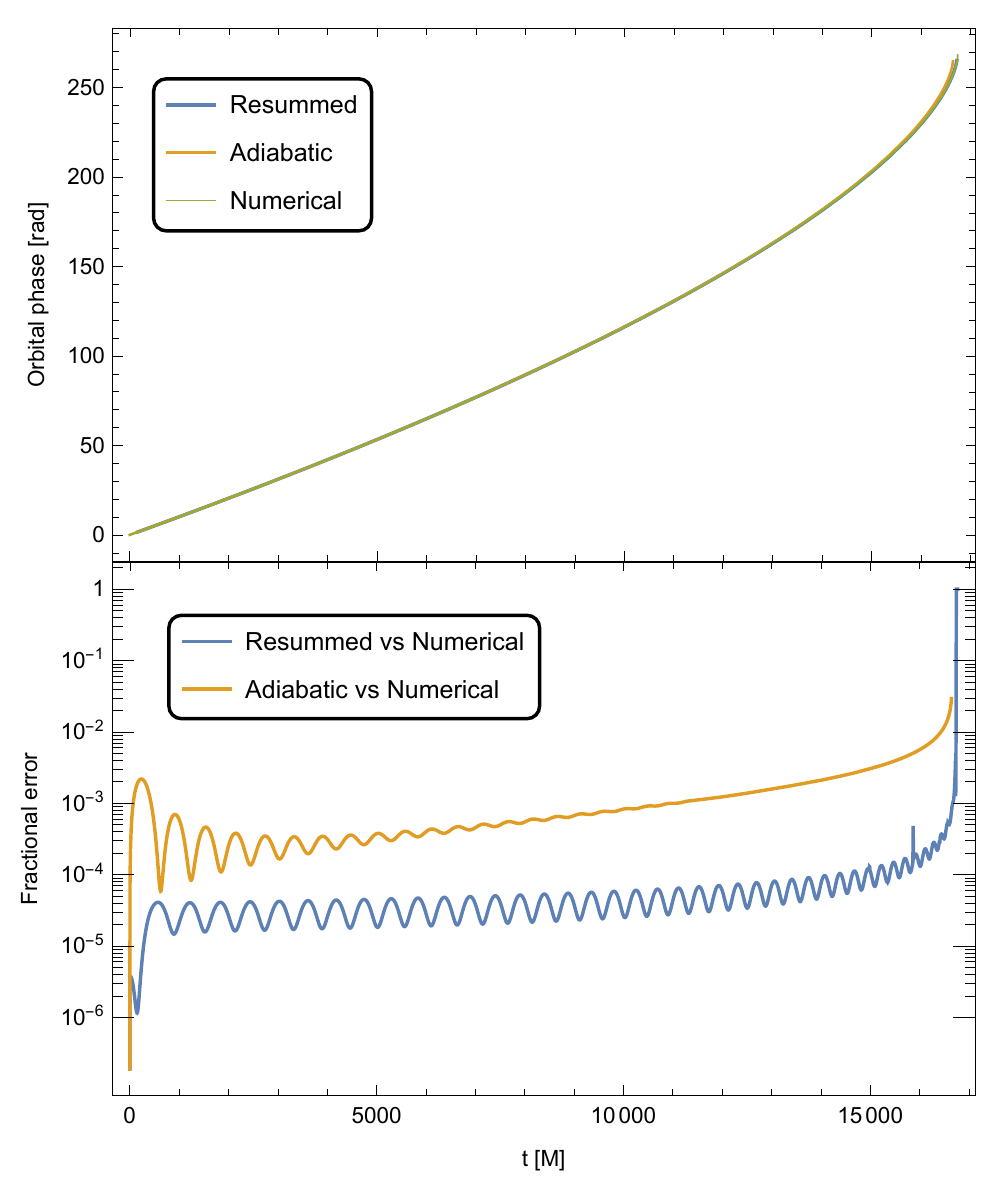}
    \includegraphics[width=0.48\columnwidth,trim = 0 0.1cm 0 0.2cm,clip]{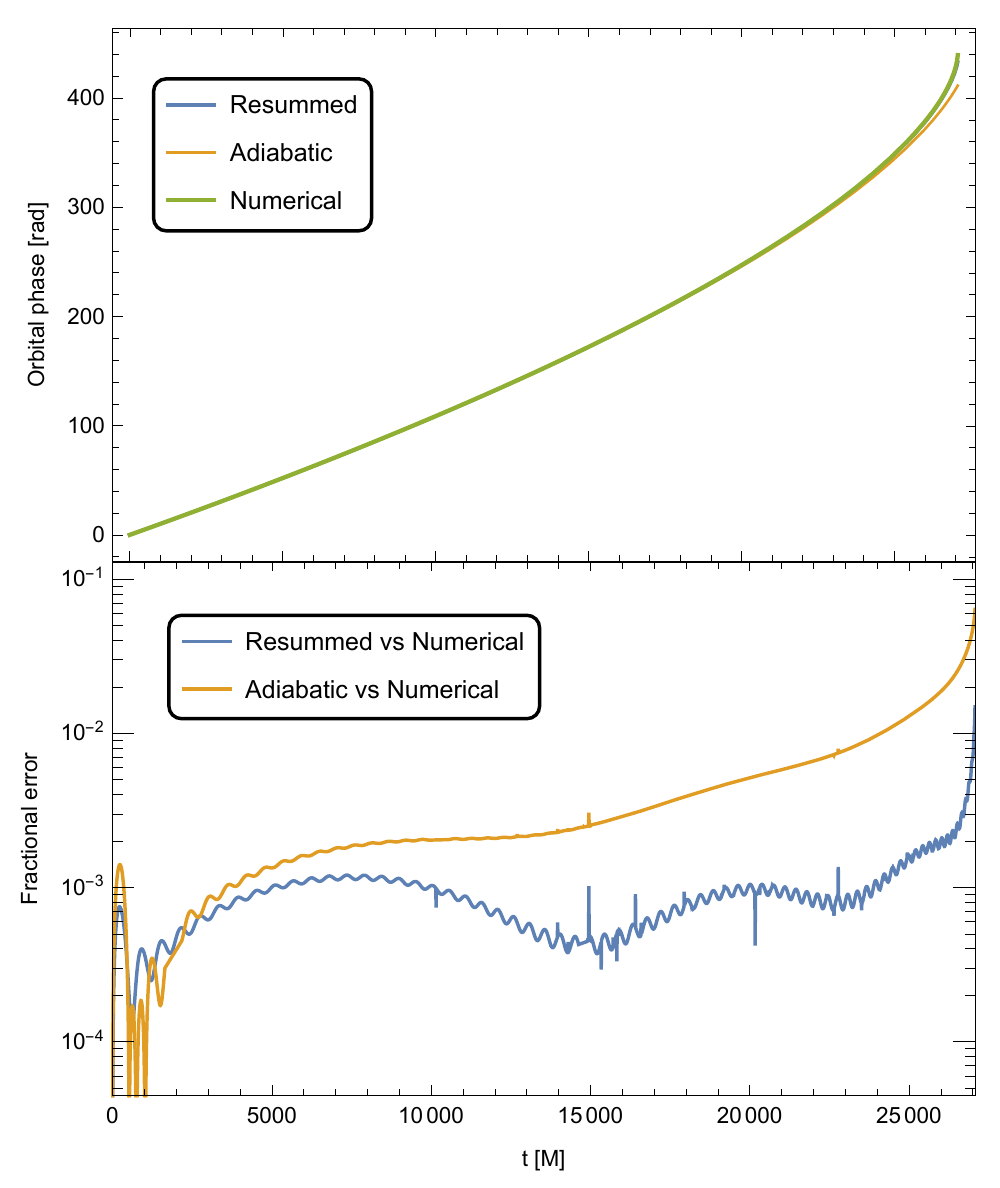}
    \label{orbit_comparison}
    \caption{{\bf Left Column}: Compact binary with equal component mass and anti-aligned initial spin vectors. {\bf Right Column}: Compact binary with component mass ratio $m_1:m_2=4$ and misaligned initial spin vectors. The initial spin configurations are given by (\ref{mratio1_init_spin}) and (\ref{mratio4_init_spin}), respectively.\\
     The first and third rows are the plots for physical values: the orbital radius and phase versus time with initial data given in (\ref{orbital_data}), respectively. The analytical renormalization group resummed solutions are plotted in \textcolor{NavyBlue}{\bf  blue}, the adiabatic solutions are  in \textcolor{Orange}{\bf orange}, and the numerical solutions to the leading order spin-radiation equations of motion are in \textcolor{LimeGreen}{\bf green}. Below each physical plot the fractional errors are shown, comparing the numerical solutions with analytical resummed solutions  in \textcolor{NavyBlue}{\bf blue} and the adiabatic solutions in \textcolor{Orange}{\bf orange}.}
\end{figure}

For initial spin vectors we consider the compact components maximally rotating, meaning the dimensionless spin parameter $\chi \sim 1$ where for each spin $|\bm{S}_a| = \chi_a m_a^2$, with $\chi_{\textrm{max}} =1$ for black holes.
In Fig.~\ref{orbit_comparison} we compare the resummed solutions to the orbital equations of motion with the numerical and adiabatic solutions \cite{Kidder:1992fr} for two different choices of mass ratio and spins. For the left column, we choose an equal mass binary and anti-aligned spin initial configuration:
\begin{align}
    \label{mratio1_init_spin}
     \frac{\bm{S}_1(0)}{m_1^2} = \cos{70^{\circ}}  \hat{\bm{n}} + \cos{60^{\circ}} \hat{\bm{\lambda}} + \cos{140^{\circ}} \hat{\bm{l}}, \qquad \frac{\bm{S}_2(0)}{m_2^2} = \frac{\cos{70^{\circ}}\cos{50^{\circ}}}{\cos{140^{\circ}}} \hat{\bm{n}}  + \frac{\cos{60^{\circ}}\cos{50^{\circ}}}{\cos{140^{\circ}}}  \hat{\bm{\lambda}} + \cos{50^{\circ}} \hat{\bm{l}}.
\end{align}
The physical interpretation for the angle of $140^{\circ}$ and $50^{\circ}$ is the angle between the spin vectors and the orbital angular momentum $\bm{L}$. (At linear
order in spin, equal mass systems satisfy the spin-orbit resonance orientations \cite{Schnittman:2004vq}.) In the right column, we choose a moderate mass ratio ($m_1:m_2 = 4$), with a randomly chosen initial spin configuration:
\begin{align}
    \label{mratio4_init_spin}
     \frac{\bm{S}_1(0)}{m_1^2} = 0.4  \hat{\bm{n}} - 0.7 \hat{\bm{\lambda}} + 0.5 \hat{\bm{l}}, \qquad \frac{\bm{S}_2(0)}{m_2^2} = 0.9 \hat{\bm{n}}  + 0.1  \hat{\bm{\lambda}} - 0.4 \hat{\bm{l}}.
\end{align}
Specifically, the plots show the orbital radius $r(t)$ and orbital phase $\phi(t)$ for resummed, adiabatic, and numerical solutions to the binary equations of motion. Below each plot of the physical solutions are the fractional errors comparing the numerical results to resummed and adiabatic ones. From these plots, we can see the DRG methods are more accurate compared to the adiabatic solutions, with roughly an order of magnitude improvement in calculating the accumulated orbital phase over most of the inspiral. 

We can see the importance of using the DRG method increases as we include higher-order corrections by comparing Fig.~\ref{orbit_comparison} to the results in Ref.~\cite{Galley:2016zee}. In that paper, the authors included the 0PN (i.e. Newtonian) contribution and the leading order radiation reaction term. As can be seen by looking at Fig.~1 of that paper, the DRG and adiabatic results give the same order relative errors.\footnote{Note that the authors of Ref.~\cite{Galley:2016zee} show how to obtain the result including 1PN contributions, but did not provide any numerical results. They also did the ``two-loop" contribution, which includes $\mathcal{O}(v^{10})$ corrections. Including these, the DRG method shows roughly an order of magnitude improvement compared to the adiabatic solution, as can be seen in Fig.~2 of that paper.} When including the 1.5PN spin contribution as we did here, there is an order of magnitude improvement, as shown in Fig.~\ref{orbit_comparison}.

\begin{figure}
    \includegraphics[width=0.48\columnwidth,trim = 0 0.5cm 0 -1.2cm]{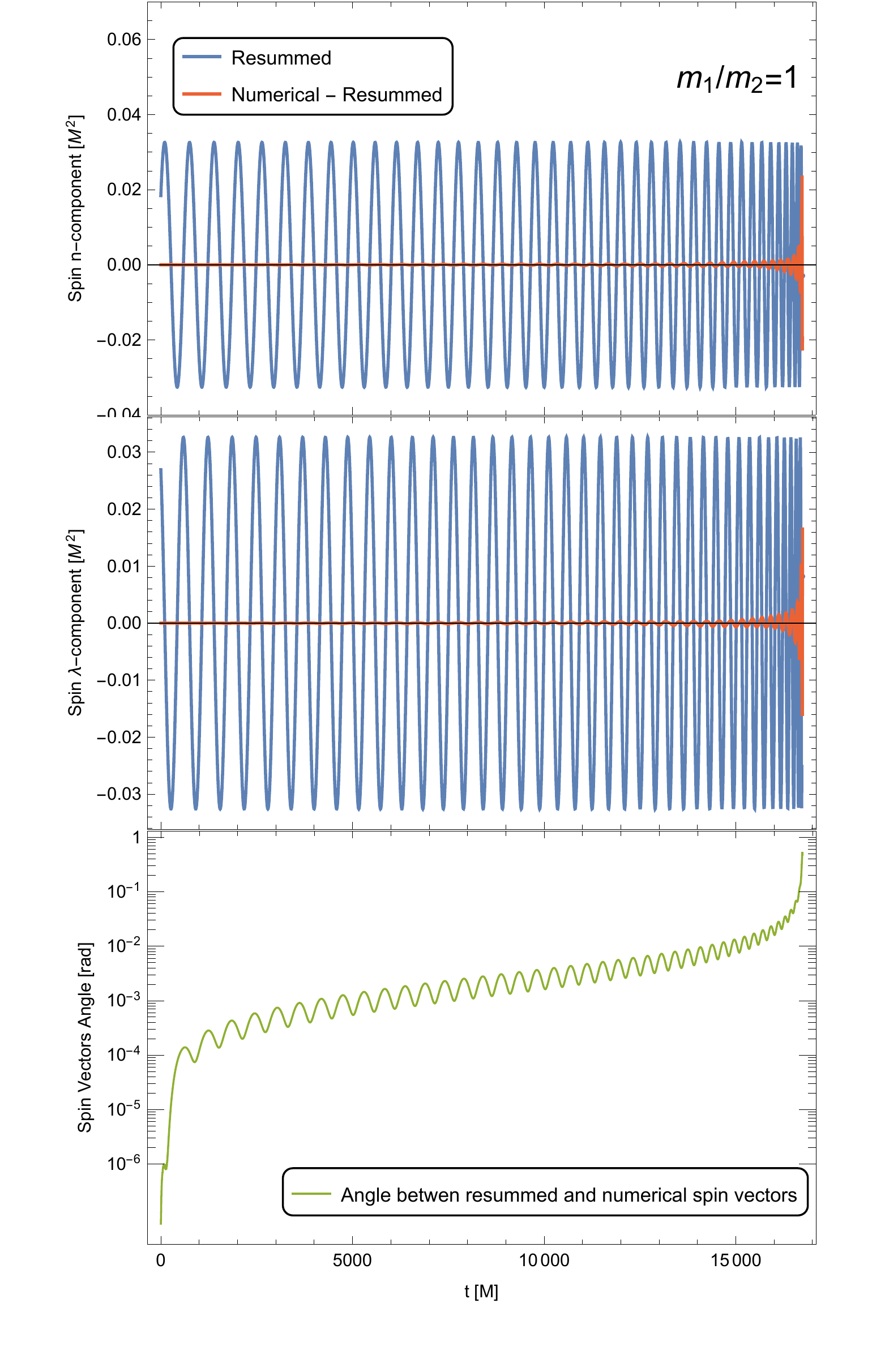}
    \includegraphics[width=0.485\columnwidth,trim = 0 0cm 0 1.5cm]{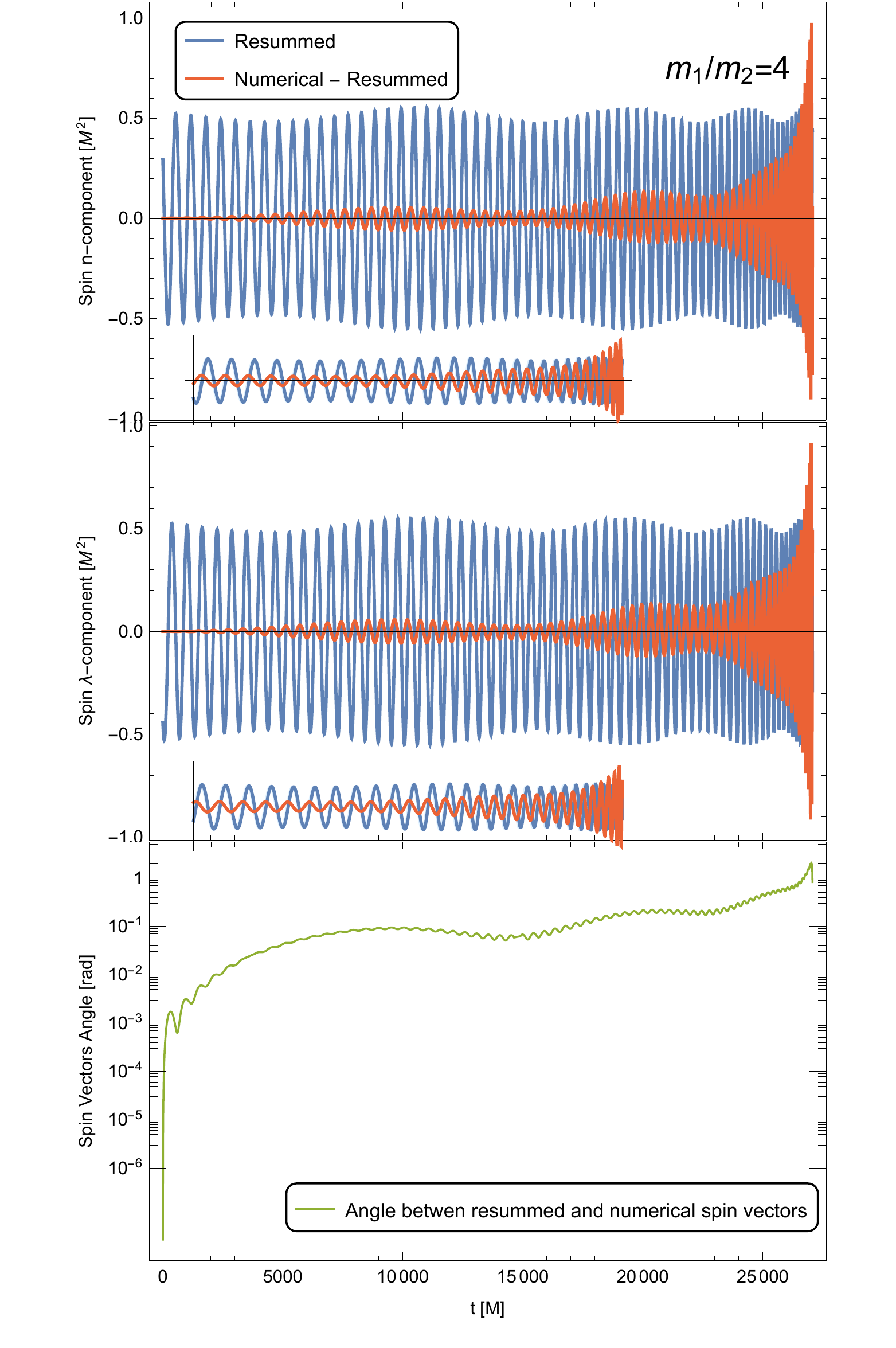}
    \includegraphics[width=0.48\columnwidth]{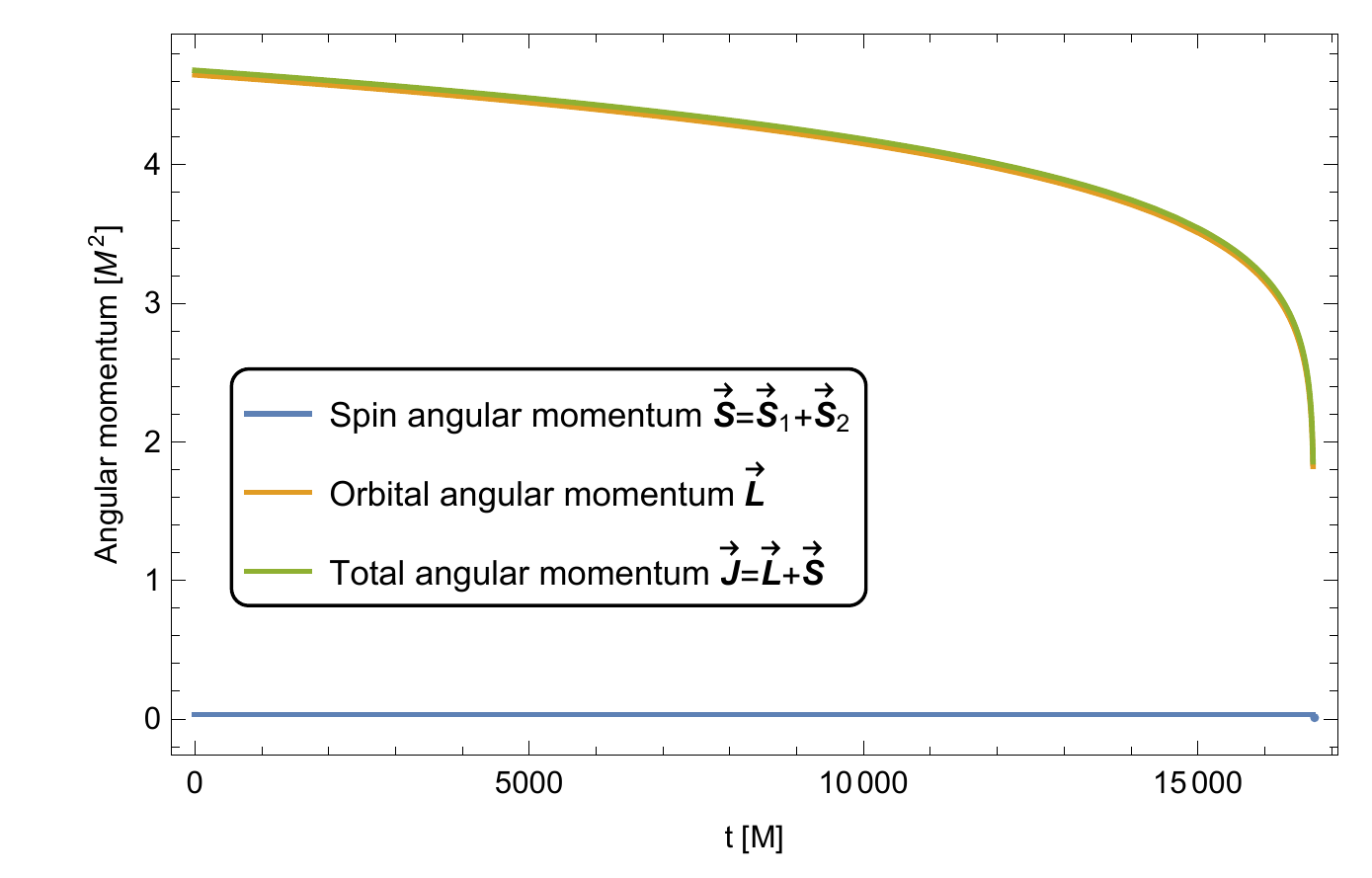}
    \includegraphics[width=0.48\columnwidth]{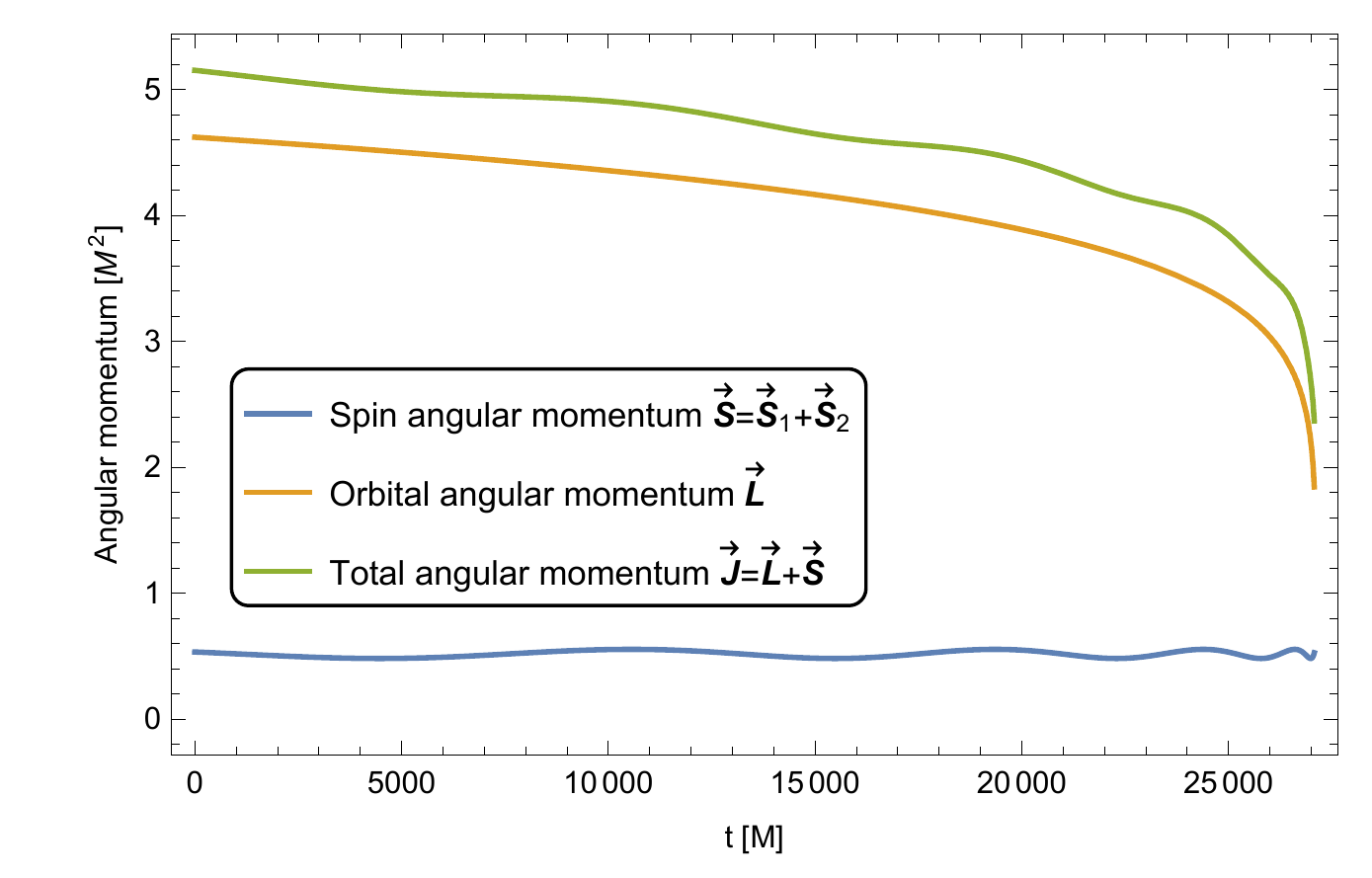}
    \label{spin_comparison}
    \caption{{\bf Left Column}: Compact binary with equal component mass and anti-aligned initial spin vectors. {\bf Right Column}: Compact binary with component mass ratio $m_1:m_2=4$ and misaligned initial spin vectors. The initial spin configurations are given by (\ref{mratio1_init_spin}) and (\ref{mratio4_init_spin}), respectively.\\
    In the first two rows from top down, the resummed solutions are in \textcolor{NavyBlue}{\bf blue}, for the corresponding spin vectors in $n$-component and $\lambda$-component. The difference of the resummed results from the numerical ones is shown in \textcolor{Red}{\bf red}. The lower inset on the right zooms in on the spin precession for the last 1/4 part of the inspiral. The third row shows the angle between the spin vector derived from the resummed solutions and the numerical solutions. In the last row, the instantaneous change of spin, orbital, and total angular momenta are shown.}
\end{figure}

We compare the resummed solutions of the spin precession equations with the numerical solutions to (\ref{spin_prec_decomp}) in Fig.~\ref{spin_comparison}. 
The two columns have the same choices for mass ratio and spin configurations as in Fig.~\ref{orbit_comparison}.  
In the top two panels, we plot the resummed solutions to the $\bm{n}$- and $\bm{\lambda}$-components, respectively, for the total spin vector (in blue) and the difference between the resummed and numerical solution (in red). 
We also include an inset plot of the spin precession for the last quarter of the inspiral to illustrate the phase difference. That the error accumulated from the resummed results of the spin precession becomes significant is the consequence of the Post-Newtonian method breaking down for large velocities during the later portion of the inspiral. We expect better accuracy when spin-spin effects and higher PN order terms are incorporated. In the third panel, we plot the angle between the spin vector results from the resummed and numerical solutions. 

With the inclusion of radiation reaction, the total angular momentum changes direction and magnitude. In the  bottom row of Fig.~\ref{spin_comparison}, we show the angular momenta changing throughout the inspiral. The  equal mass binary shown in the left panel has a fixed total spin magnitude due to the symmetric form in (\ref{spinprec}). Both binaries exhibit a rapid loss of orbital and total angular momenta at the end of the inspiral in sync with the drop of the orbital radius in Fig~\ref{orbit_comparison}.  

\begin{SCfigure}
  \centering
    \caption{C++ runtime comparison between numerical integration and DRG resummed substitution. The count in the x-axis stands for the total choices of initial conditions in a particular run, and green numbers below the plot points are the average steps taken per run. The blue dots show the total computation time for the numerical integration solutions and the orange squares shows the time for the DRG resummed results substitutions.\\\\}
    \includegraphics[width=0.75\columnwidth]{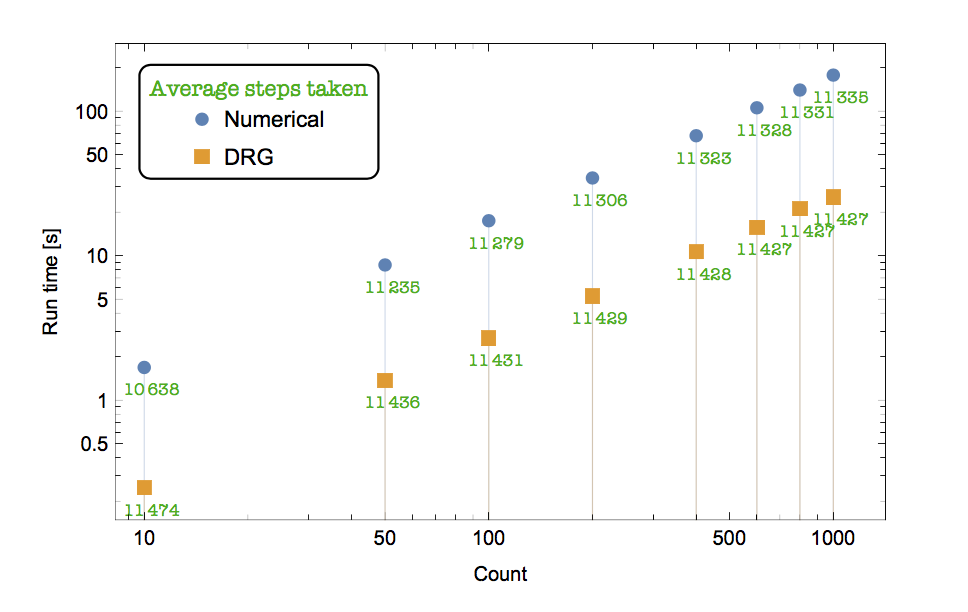}
    \label{cpp}
\end{SCfigure}

In Fig.~\ref{cpp} we give a rough comparison of the computational runtime improvement of the DRG methods. The numerical solution for the equations of motion and spin precession was calculated in C++ implementing the ODEINT library \cite{ahnert2011odeint}. We adopt the Dormand-Prince algorithm at fifth order with adaptive step sizes and control the tolerance error to be consistent with the theoretical resummed solution errors. Fig.~\ref{cpp} shows the runtime of the numerical and DRG methods solving the same sets of initial conditions, changing the binary mass ratio Count times in each run. In order to try to have a meaningful comparison, we manipulate the average steps taken per run for the DRG methods to have similar output lengths (i.e., number of time steps for the solution) with the numerical integration. For example, in a total of 50 runs, the numerical integration takes 10 seconds and averagely 11235 steps per run, while the DRG method takes about 1 second and averagely 11438 steps per run. As can be seen, the DRG method is an order of magnitude faster than the numerical solution.

\section{Conclusion}
\label{section_Conclusion}
Using the dynamical renormalization group formalism, we have solved the spinning binary dynamics including the 2.5PN radiation reaction and the leading order spin-orbit effects throughout the inspiral. The solution is obtained by the resummation of the secularly growing perturbations to the physical parameters including orbital radius, angular frequency, orbital phase, and spin precession phases. We solved the resummed solutions to the equations of motion and spin precession equations in a moving triad frame at any time instant. Renormalized parameters defined to describe the resummed solutions are determined using the renormalization group equations and can be written in terms of conserved identities.

The solutions are applicable to arbitrary initial configurations and do not dependent on any specific spin orientations. The comparison of numerical solutions and our analytic solutions shows greater accuracy than the adiabatic solutions and a sizable improvement in computation time compared to the numerical solutions. The use of the DRG method is more important for spinning BHs than for the non-spinning case. However,  there is further improvements that can be made. The spin component comparison is not ideal, as shown in Fig.~\ref{spin_comparison} with  increasing phase differences. When initial spins are relatively large compare to orbital angular momentum, the discrepancy grows very fast in the early part of the inspiral. This is due to the beginning of the breakdown of the PN expansion. We hope to fix this issue and enhance the accuracy by the inclusion of spin-spin effects and higher-order PN terms into the formulation in the future works \cite{Zixin}.    

\section{Acknowledgments}

We thank Ira Rothstein and Chad Galley for useful discussions. We also thank Shana Li for useful discussions.  ZY and AKL were supported in part by the National Science Foundation under Grant No. PHY-1820760.

\appendix 

\section{Orbital Equations of Motion Resummation Solutions}
\label{appendix_orb_solutions}

We start by investigating the quasi-circular background orbit solution of the conservative spinning binary. In this case, the radius and the orbital angular frequency are constants apart from small non-secular perturbations induced by the presence of spins. The constant radius $R_B$ and orbital frequency $\Omega_B$ satisfy
\begin{equation}
     \Omega^2_B = \frac{M}{R^3_B} - \frac{\Omega_B}{R_B^3}\big(5S_l + 3\Delta\Sigma_l\big) + \mathcal{O}(S^2).
 \end{equation} 
at  linear order in spin. Setting the spins to zero, the relation between $R_B$ and $\Omega_B$ reduces to the usual Newtonian circular motion equation. To solve for $\Omega_B$ with a given $R_B$, we can either solve the quadratic equation above or substitute for $\Omega_B$ iteratively. The analytic solutions to the dynamics of quasi-circular conservative spinning binary systems have been studied \cite{Blanchet:2011zv,marsat2013next}.

\subsection{Perturbations of quasi-circular orbits}
Next we describe the deviation of the quasi-circular background orbit as a result of the leading order radiation reaction and linear spin-orbit effects by isolating the perturbative corrections $r(t) = R_B + \delta r(t) + \delta r_S (t)$ and $\omega(t) = \Omega_B + \delta \omega(t) + \delta \omega_S (t)$. The first time-dependent terms $\delta r(t)$ and $\delta \omega(t)$ are the perturbation that arise due to the 2.5PN radiation reaction force without the spin at a given time $t$, which are given by \cite{Galley:2016zee}
\begin{subequations}
\label{delta_RR}
    \begin{align}
            \delta r(t) &= -\frac{64\nu}{5}R^6_B\Omega_B^6 (t-t_0) +A_B \sin{(\Omega_B(t-t_0)+\Phi_B)}, \label{delta_r} \\
            \delta \omega(t) &= \frac{96\nu}{5}R^5_B\Omega_B^7 (t-t_0) -\frac{2\Omega_B A_B}{R_B} \sin{(\Omega_B(t-t_0)+\Phi_B)},\label{delta_omega}
    \end{align}
\end{subequations}
 with bare parameters $\{R_B,\Omega_B,A_B,\Phi_B\}$, and $\delta r \sim v^5 R_B, \delta \omega \sim  v^6/R_B$ at the initial time $t_0$. On the other hand, the terms due to the interaction of 1.5PN spin effects and the 2.5PN radiation reaction start with the power counting of $\delta r_S \sim S v^4 /R_B$ and $\delta \omega_S \sim S v^5/R_B^3$. Expanding the equations of motion (\ref{EoM_r}) and (\ref{EoM_omega}) to $\mathcal{O}(S v^{6})$ gives
\begin{subequations}
\begin{align}
    \delta \ddot{r}_S(t) - 2R_B\Omega_B \delta \omega_{S}(t) -3 \Omega_B^2\delta r_S(t) &=\frac{\delta\omega(t) }{R_B^2}(5S_l + 3\Delta \Sigma_l), \label{nhat_eq}  \\
    R_B \delta \dot{\omega}_S(t) + 2\Omega_B \delta \dot{r}_S(t) &= - \left(\frac{2S_l}{R_B^3} \delta\dot{r}(t) + \big(88S_l+\frac{264}{5}\Delta \Sigma_l\big)\nu R_B^3 \Omega_B^6  \right), \label{lambda_eq}
\end{align}
\end{subequations}
with $\delta r(t)$ and $\delta \omega(t)$ the values given in (\ref{delta_r}) and (\ref{delta_omega}). Integrating (\ref{lambda_eq}) with respect to time, solving for $\delta\omega_S$, and substituting back into (\ref{nhat_eq}) gives the differential equation for $\delta r_S$,
\begin{align}
    \delta\ddot{r}_S(t) + \Omega_B^2\delta r_S(t) = & -\Big(\frac{144}{5}S_l + 48\Delta\Sigma_l \Big)\nu R_B^3 \Omega_B^7 (t-t_0) \nonumber \\
    & - (14S_l +6\Delta\Sigma_l)\frac{\Omega_B A_B}{R_B^3} \sin{(\Omega_B(t-t_0)+\Phi_B)} .\label{r_DE}
\end{align}
The differential equation has a solution of the form
\begin{subequations}
\label{del_pert_S}
\begin{align}
    \delta r_S(t) =& -\Big(\frac{144}{5}S_l + 48\Delta\Sigma_l \Big)\nu R_B^3 \Omega_B^5 (t-t_0) \nonumber\\
    &+ \frac{\big(7S_l +3\Delta\Sigma_l\big)}{2\Omega_B R_B^3}A_B \Big[2\Omega_B(t-t_0)\cos{(\Omega_B(t-t_0)+\Phi_B)} - \sin{(\Omega_B(t-t_0)+\Phi_B)} \Big] \nonumber\\
    &+ A^S_B \cos{(\Omega_B(t-t_0)+\Phi_B)}, \label{del_r}
\end{align}
where $A^S_B \sim S v^4/ R_B$ is a bare parameter in the general solution to the homogeneous equation of (\ref{r_DE}), to be determined by initial conditions. While $e_B= A_B / R_B$ is the small orbital eccentricity of order $\mathcal{O}(v^5)$ induced by the radiation reaction force, the interaction between the spin and radiation reaction lead to a smaller eccentricity $e^S_B = A^S_B/R_B \sim \mathcal{O}(Sv^4)$. The spin-radiation eccentricity deforms the circular orbit out-of-phase relative to the radiation eccentricity, with a fixed phase difference. 

As a result, the angular frequency perturbation $\delta\omega_S(t)$ and its time integration $\delta\Phi_S(t)$ are given by
\begin{align}
    \delta\omega_S(t) = &\Big(-\frac{24}{5}S_l + \frac{216}{5}\Delta \Sigma_l \Big)\nu R^2_B \Omega_B^6 (t-t_0) + \big(5S_l + 3\Delta \Sigma_l \big)\frac{A_B}{R_B^4}\sin{(\Omega_B(t-t_0)+\Phi_B)} \nonumber \\
    &- \big(14S_l + 6\Delta \Sigma_l \big)\frac{A_B}{R_B^4}\Omega_B(t-t_0)\cos{(\Omega_B(t-t_0)+\Phi_B)} - \frac{2A^S_B \Omega_B}{R_B} \cos{(\Omega_B(t-t_0)+\Phi_B)},\label{del_omega} \\
    \delta\Phi_S(t) = &\Big(-\frac{12}{5}S_l + \frac{108}{5}\Delta \Sigma_l \Big)\nu R^2_B \Omega_B^6 (t-t_0)^2 - \Big(19S_l + 9\Delta \Sigma_l \Big)\frac{A_B}{\Omega_B R_B^4}\cos{(\Omega_B(t-t_0)+\Phi_B)} \nonumber \\
    & - \Big(14S_l + 6\Delta \Sigma_l \Big)\frac{A_B}{R_B^4}(t-t_0)\sin{(\Omega_B(t-t_0)+\Phi_B)} - \frac{2A^S_B}{R_B} \sin{(\Omega_B(t-t_0)+\Phi_B)}. \label{del_phi}
\end{align}
\end{subequations}
The perturbation $\delta\Phi_S(t)$, of order $\mathcal{O}(Sv^4)$ to the angle $\phi(t)$, is the analog of the orbital phase in planar motion of non-spinning systems. Though it is no longer a physical angle now that the orbital plane precesses due to the spins, it is a combination of the Euler angles, defined in a later section, essential to the time evolution of the moving frame of reference. 

We split the perturbation terms into the non-secular terms that remain small permanently, and the secular ones that grows with time. As time progresses, the secular terms gradually become dominant and break down the PN power counting, therefore they need to be resummed. 

\subsection{Renormalization}

The full set of bare solutions to the orbit motion including linear spin-orbit terms and 2.5PN Burke-Thorne terms is given by
\begin{subequations}
\label{delta_r_omega}
\begin{align}
    r(t) &= R_B + \delta r(t) + \delta r_S(t),\\
    \omega(t) &= \Omega_B + \delta \omega(t) + \delta \omega_S(t),\\
    \phi(t) &= \Phi_B + \delta \Phi(t) + \delta \Phi_S(t),
\end{align}
\end{subequations}
with the corresponding perturbations in (\ref{delta_RR}) and (\ref{del_pert_S}). We renormalize these terms by removing the $t_0$ dependence with the introduction of counter-terms for the bare parameters.  The $\mathcal{O}(v^5)$ terms were renormalized in Ref.~\cite{Galley:2016zee}.
Thanks to the newly added $\mathcal{O}(Sv^4)$ perturbations, the bare parameters have to include higher order counter-terms, which means
\begin{subequations}
\begin{align}
    R_B (t_0) &= R_R(\tau) + \delta_R^{v^5}(\tau,t_0) + \delta_R^{S}(\tau,t_0),\\
    \Omega_B (t_0) &= \Omega_R(\tau) + \delta_\Omega^{v^5}(\tau,t_0) + \delta_\Omega^{S}(\tau,t_0),\\
    \Phi_B (t_0) &= \Phi_R(\tau) + \delta_\Phi^{v^5}(\tau,t_0) + \delta_\Phi^{S}(\tau,t_0),\\
    A^S_B (t_0) &= A^S_R(\tau) + \delta_A^{S}(\tau,t_0). 
\end{align}
\end{subequations}

In terms of the renormalized initial parameters and the renormalization scale $t-t_0 = (t-\tau) + (\tau-t_0)$, the spin-orbit result becomes
\begin{subequations}
\begin{align}
    r(t) =& R_R + \delta_R^{S} - \frac{64\nu}{5}R_R^6 \Omega_R^6 (t-\tau) +A_R \sin{(\Omega_R(t-\tau)+\Phi_R)} \nonumber \\
    &-\left(\frac{144}{5}S_l + 48\Delta\Sigma_l \right)\nu R_R^3 \Omega_R^5 (t-\tau)-\left(\frac{144}{5}S_l + 48\Delta\Sigma_l \right)\nu R_R^3 \Omega_R^5 (\tau-t_0) \nonumber\\
    &+ \frac{\big(7S_l +3\Delta\Sigma_l\big)}{2\Omega_R R_R^3}A_R \Big[2\Omega_R(t-\tau)\cos{(\Omega_R(t-\tau)+\Phi_R)} + 2\Omega_R(\tau-t_0)\cos{(\Omega_R(t-\tau)+\Phi_R)} \nonumber\\
    &\qquad\qquad\qquad\qquad\qquad- \sin{(\Omega_R(t-\tau)+\Phi_R)} \Big]\nonumber\\
    &+ A^S_R \cos{(\Omega_R(t-\tau)+\Phi_R)} + \delta_A^{S} \cos{(\Omega_R(t-\tau)+\Phi_R)},\\\nonumber\\
    \omega(t) =& \Omega_R + \delta_\Omega^{S} + \frac{96\nu}{5}R_R^5 \Omega_R^7 (t-\tau) - \frac{2\Omega_R A_R}{R_R} \sin{(\Omega_R(t-\tau)+\Phi_R)} \nonumber \\
    &+\Big(-\frac{24}{5}S_l + \frac{216}{5}\Delta \Sigma_l \Big)\nu R^2_R \Omega_R^6 (t-\tau) +\Big(-\frac{24}{5}S_l + \frac{216}{5}\Delta \Sigma_l \Big)\nu R^2_R \Omega_R^6 (\tau-t_0) \nonumber \\
    &+ \Big(5S_l + 3\Delta \Sigma_l \Big)\frac{A_R}{R_R^4}\sin{(\Omega_R(t-\tau)+\Phi_R)}- \Big(14S_l + 6\Delta \Sigma_l \Big)\frac{A_R}{R_R^4}\Omega_R(t-\tau)\cos{(\Omega_R(t-\tau)+\Phi_R)}\nonumber\\
    &- \Big(14S_l + 6\Delta \Sigma_l \Big)\frac{A_R}{R_R^4}\Omega_R(\tau-t_0)\cos{(\Omega_R(t-\tau)+\Phi_R)}\nonumber\\
    &- \frac{2A^S_R \Omega_R}{R_R} \cos{(\Omega_R(t-\tau)+\Phi_R)} - \frac{2\delta_A^{S} \Omega_R}{R_R} \cos{(\Omega_R(t-\tau)+\Phi_R)},\\\nonumber\\
    \phi(t) =& \Phi_R + \delta_\Phi^{S} + (t-\tau)\Omega_R + (t-\tau)\delta_\Omega^{S} + (\tau-t_0)\delta_\Omega^{S}+ \frac{48\nu}{5}R_R^5 \Omega_R^7 (t-\tau)^2 + \frac{2A_R}{R_R} \cos{(\Omega_R(t-\tau)+\Phi_R)} \nonumber \\
    &- \Big(\frac{12}{5}S_l - \frac{108}{5}\Delta \Sigma_l \Big)\nu R^2_R \Omega_R^6 (t-\tau)^2 + \Big(-\frac{24}{5}S_l + \frac{216}{5}\Delta \Sigma_l \Big)\nu R^2_R \Omega_R^6 (t-\tau)(\tau-t_0) \nonumber\\
    &+ \Big(-\frac{12}{5}S_l + \frac{108}{5}\Delta \Sigma_l \Big)\nu R^2_R \Omega_R^6 (\tau-t_0)^2 - \Big(19S_l + 9\Delta \Sigma_l \Big)\frac{A_R}{\Omega_R R_R^4}\cos{(\Omega_R(t-\tau)+\Phi_R)} \nonumber \\
    & - \Big(14S_l + 6\Delta \Sigma_l \Big)\frac{A_R}{R_R^4}(t-\tau)\sin{(\Omega_R(t-\tau)+\Phi_R)} - \Big(14S_l + 6\Delta \Sigma_l \Big)\frac{A_R}{R_R^4}(\tau-t_0)\sin{(\Omega_R(t-\tau)+\Phi_R)}\nonumber\\
    & - \frac{2A^S_R}{R_R} \sin{(\Omega_R(t-\tau)+\Phi_R)} - \frac{2\delta_A^{S}}{R_R} \sin{(\Omega_R(t-\tau)+\Phi_R)}. 
\end{align}
\end{subequations}
By observation we can write down the counter-terms that cancel the $(\tau-t_0)$ terms completely as
\begin{subequations}
\label{counter_S}
\begin{align}
     \delta^{S}_R(\tau,t_0) =& \left(\frac{144}{5}S_l + 48\Delta\Sigma_l \right)\nu R_R^3 \Omega_R^5 (\tau-t_0),\\
     \delta^{S}_\Omega(\tau,t_0) =& \Big(\frac{24}{5}S_l - \frac{216}{5}\Delta \Sigma_l \Big)\nu R^2_R \Omega_R^6 (\tau-t_0),\\
     \delta^{S}_\Phi(\tau,t_0) =&  \Big(-\frac{12}{5}S_l + \frac{108}{5}\Delta \Sigma_l \Big)\nu R^2_R \Omega_R^6 (\tau-t_0)^2,\\
     \delta^{S}_A(\tau,t_0) =& - \Big(7S_l + 3\Delta \Sigma_l \Big)\frac{A_R}{R_R^3}(\tau-t_0).
 \end{align} 
\end{subequations}
Choosing the arbitrary renormalization scale to be $\tau = t_0$, the equations of motion is now described by the renormalized quantities $\{R_R, \Omega_R, \Phi_R, A_R, A_R^S \}$ as 
\begin{subequations}
    \label{orbit_resummed}
\begin{align}
    r(t) =& R_R(t) + \Bigg(1 - \frac{\big(7S_l +3\Delta\Sigma_l\big)}{2\Omega_R(t) R_R^3(t)}\Bigg)A_R(t) \sin{\Phi_R(t)} + A^S_R(t)\cos{\Phi_R(t)},\\
    \omega(t) =& \Omega_R(t) - \frac{2\Omega_R(t)A_R(t)}{R_R(t)}\Bigg(1 - \frac{\big(5S_l +3\Delta\Sigma_l\big)}{2\Omega_R(t) R_R^3(t)}\Bigg)\sin{\Phi_R(t)} - \frac{2A^S_R \Omega_R(t)}{R_R(t)} \cos{\Phi_R(t)},\\
    \phi(t) =& \Phi_R(t) + \frac{2A_R(t)}{R_R(t)}\Bigg(1 - \frac{\big(19S_l +9\Delta\Sigma_l\big)}{2\Omega_R(t) R_R^3(t)}\Bigg)\cos{\Phi_R(t)} - \frac{2A^S_R(t) }{R_R(t)} \sin{\Phi_R(t)}. \label{phi_resummed}
\end{align}
\end{subequations}
The explicit secular terms have been removed thanks to the choice of $\tau$, and the $t_0$-dependencies have been absorbed into the counter-terms. The runnings of $\{R_R, \Omega_R, \Phi_R, A_R, A_R^S \}$ and their dependence on the initial conditions are then determined by the renormalization group equations.

\subsection{Renormalization Group Solutions}      
\label{EoM_RGE}

Exploiting the fact that the bare quantities $\{R_B(t_0), \Omega_B(t_0), \Phi_B(t_0), A^S_B(t_0) \}$ are independent of the arbitrary scale $\tau$, we can write down the renormalization group equations for the renormalized quantities $\{R_R(t), \Omega_R(t), \Phi_R(t), A^S_R(t)\}$ as
\begin{subequations}
    \label{RGE}
\begin{align}
    \frac{\ud R_R}{\ud \tau} =& -\frac{64\nu}{5}R_R^6(\tau)\Omega_R^6(\tau) - \left(\frac{144}{5}S_l + 48\Delta\Sigma_l \right)\nu R_R^3(\tau) \Omega_R^5(\tau), \label{RGE_R} \\
    \frac{\ud \Omega_R}{\ud \tau} =& \frac{96\nu}{5}R_R^5(\tau)\Omega_R^7(\tau) - \Big(\frac{24}{5}S_l - \frac{216}{5}\Delta \Sigma_l \Big)\nu R^2_R(\tau) \Omega_R^6(\tau), \label{RGE_Omega}\\
    \frac{\ud \Phi_R}{\ud \tau} =& \Omega_R(\tau), \\
    \frac{\ud A^S_R}{\ud \tau} =& \Big(7S_l + 3\Delta \Sigma_l \Big)\frac{A_R}{R_R^3}.\label{RGE_A_S}
\end{align}
\end{subequations}
The right-hand sides of the RG equations, which are called beta functions, includes more iterative time-derivative terms that are of higher orders starting from $\mathcal{O}(v^{11})$ and $\mathcal{O}(S^2) $. The RG solutions to $\Omega_R$, $\Phi_R$ and $A^S_R$ in terms of $R_R$ and the initial conditions are
\begin{subequations}
\label{Renorm_Sol}
\begin{align}
    \Omega_R(t) =&\left[\frac{M^{1/2}}{R_R^{3/2}(t)} - \left(\frac{5S_l + 3\Delta\Sigma_l}{2R_R^3(t)} \right) \right] =\left[\frac{M}{R_R^3(t)} - \sqrt{\frac{M}{R_R^3(t)}}\left(\frac{5S_l + 3\Delta\Sigma_l}{R_R^3(t)} \right) \right]^{\frac{1}{2}}+ \mathcal{O}(S^2),\\\nonumber\\
    \Phi_R(t) =& \Phi_R(t_i) + \frac{1}{32M^{5/2}\nu}\Big[R_R^{5/2}(t_i) - R_R^{5/2}(t) \Big] + \frac{5\Big(41S_l + 15\Delta \Sigma_l \Big)}{256M^3\nu} \Big[R_R(t_i) - R_R(t) \Big] \nonumber\\
    &+ \frac{5\mathscr{S}^{2/3}\Big(41S_l + 15\Delta \Sigma_l \Big)}{128\sqrt{3}\nu M^{10/3}}\Bigg[\tan^{-1}\Bigg(\frac{1}{\sqrt{3}}\Big(1 + \frac{2M^{1/6}R_R(t_i)^{1/2}}{\mathscr{S}^{1/3}} \Big)\Bigg) -\tan^{-1}\Bigg(\frac{1}{\sqrt{3}}\Big(1 + \frac{2M^{1/6}R_R(t)^{1/2}}{\mathscr{S}^{1/3}} \Big)\Bigg)\Bigg]\nonumber\\
    &+ \frac{5\mathscr{S}^{2/3}\Big(41S_l + 15\Delta \Sigma_l \Big)}{768\nu M^{10/3}}\Bigg[\ln\Bigg(\frac{\Big(\mathscr{S}^{1/3}-M^{1/6}R_R(t_i)^{1/2}\Big)^2}{\mathscr{S}^{2/3} + \mathscr{S}^{1/3}M^{1/6}R_R(t_i)^{1/2} + M^{1/3}R_R(t_i)} \Bigg) \nonumber\\
    &\qquad\qquad\qquad\qquad\qquad\qquad\qquad\qquad - \ln\Bigg(\frac{\Big(\mathscr{S}^{1/3}-M^{1/6}R_R(t)^{1/2}\Big)^2}{\mathscr{S}^{2/3} + \mathscr{S}^{1/3}M^{1/6}R_R(t)^{1/2} + M^{1/3}R_R(t)} \Bigg)\Bigg],\\\nonumber\\
    A^S_R(t) =& A^S_R(t_i) + \frac{5A_R}{64\nu M^3}\Big(7S_l + 3\Delta \Sigma_l \Big) \Big[R_R(t_i) - R_R(t) \Big]\nonumber\\
    &+\frac{5A_R\mathscr{S}^{2/3}\Big(7S_l + 3\Delta \Sigma_l \Big)}{32{\sqrt{3}}\nu M^{10/3}}\Bigg[\tan^{-1}\Bigg(\frac{1}{\sqrt{3}}\Big(1 + \frac{2M^{1/6}R_R(t_i)^{1/2}}{\mathscr{S}^{1/3}} \Big)\Bigg) - \tan^{-1}\Bigg(\frac{1}{\sqrt{3}}\Big(1 + \frac{2M^{1/6}R_R(t)^{1/2}}{\mathscr{S}^{1/3}} \Big)\Bigg)\Bigg]  \nonumber\\
    &+\frac{5A_R\mathscr{S}^{2/3}\Big(7S_l + 3\Delta \Sigma_l \Big)}{192\nu M^{10/3}}\Bigg[\ln\Bigg(\frac{\Big(\mathscr{S}^{1/3}-M^{1/6}R_R(t_i)^{1/2}\Big)^2}{\mathscr{S}^{2/3} + \mathscr{S}^{1/3}M^{1/6}R_R(t_i)^{1/2} + M^{1/3}R_R(t_i)} \Bigg)  \nonumber\\
    &\qquad\qquad\qquad\qquad\qquad\qquad\qquad\qquad - \ln\Bigg(\frac{\Big(\mathscr{S}^{1/3}-M^{1/6}R_R(t)^{1/2}\Big)^2}{\mathscr{S}^{2/3} + \mathscr{S}^{1/3}M^{1/6}R_R(t)^{1/2} + M^{1/3}R_R(t)} \Bigg)\Bigg],
\end{align}
\end{subequations}
where $\mathscr{S} \equiv (51S_l+21\Delta\Sigma_l)/4$ is a constant combination of the initial spins, defined for convenience. Substituting in to the radial RG equation, we find
\begin{align}
    \frac{\ud R_R}{\ud \tau} = -\frac{64\nu M^3}{5R_R^3} + \frac{16 \nu M^{5/2}}{5R_R^{9/2}}(51S_l+21\Delta\Sigma_l),
\end{align}
or, rearranging, 
\begin{align}
     \frac{R_R^{9/2}}{R^{3/2} - M^{-1/2}\mathscr{S}} \ud R_R=-\frac{64\nu M^3}{5}\ud \tau.
\end{align}
Integrating both sides gives the exact but implicit relation,
\begin{align}
\label{Renorm_Sol_R}
    -\frac{64\nu M^3}{5}(t-t_i) =& \frac{1}{4}\big(R_R(t)^4 - R_R(t_i)^4 \big) + \frac{2\mathscr{S}}{5M^{1/2}}\big(R_R(t)^{5/2} - R_R(t_i)^{5/2} \big) + \frac{\mathscr{S}^2}{M}\big(R_R(t) - R_R(t_i) \big) \nonumber \\
    &+\frac{2\mathscr{S}^{8/3}}{\sqrt{3}M^{4/3}}\Bigg[\tan^{-1}\Bigg(\frac{1}{\sqrt{3}}\Big(1 + \frac{2M^{1/6}R_R(t)^{1/2}}{\mathscr{S}^{1/3}} \Big)\Bigg) -\tan^{-1}\Bigg(\frac{1}{\sqrt{3}}\Big(1 + \frac{2M^{1/6}R_R(t_i)^{1/2}}{\mathscr{S}^{1/3}} \Big)\Bigg) \Bigg] \nonumber\\
    &+\frac{\mathscr{S}^{8/3}}{3M^{4/3}}\Bigg[\ln\Bigg(\frac{\Big(\mathscr{S}^{1/3}-M^{1/6}R_R(t)^{1/2}\Big)^2}{\mathscr{S}^{2/3} + \mathscr{S}^{1/3}M^{1/6}R_R(t)^{1/2} + M^{1/3}R_R(t)} \Bigg) \nonumber\\
    &\qquad\qquad\qquad - \ln\Bigg(\frac{\Big(\mathscr{S}^{1/3}-M^{1/6}R_R(t_i)^{1/2}\Big)^2}{\mathscr{S}^{2/3} + \mathscr{S}^{1/3}M^{1/6}R_R(t_i)^{1/2} + M^{1/3}R_R(t_i)} \Bigg)  \Bigg].
\end{align}
The parameter $A_R$ is unchanged when the spin is added, and from \cite{Galley:2016zee} we learned that $A_R$ has a zero $\beta$-function at the order we are working, i.e., $A_R$ is a constant, given by initial conditions, proportional to the initial eccentricity $e_R(0) =A_R(0)/R_R(0) \sim \mathcal{O}(v^5).$

Using the relation above for $R_R(t)$, we can further simply the expressions of $\Phi_R(t)$ and $A^S_R(t)$ in terms of $R_R(t)$ and time $t$, eliminating the logarithm and the arctangent terms. Written as an invariant in time, the renormalized quantities with the leading order spin-orbit effect are
\begin{subequations}
\begin{align}
     \frac{64\nu M^3}{5}t +\frac{1}{4}R_R(t)^4 + \frac{2\mathscr{S}}{5M^{1/2}}R_R(t)^{5/2} + \frac{\mathscr{S}^2}{M}R_R(t) +\frac{2\mathscr{S}^{8/3}}{\sqrt{3}M^{4/3}}\tan^{-1}\Bigg(\frac{1}{\sqrt{3}} + \frac{2M^{1/6}R_R(t)^{1/2}}{\sqrt{3}\mathscr{S}^{1/3}}\Bigg) &\nonumber\\
    +\frac{\mathscr{S}^{8/3}}{3M^{4/3}}\ln\Bigg(\frac{\Big(\mathscr{S}^{1/3}-M^{1/6}R_R(t)^{1/2}\Big)^2}{\mathscr{S}^{2/3} + \mathscr{S}^{1/3}M^{1/6}R_R(t)^{1/2} + M^{1/3}R_R(t)} \Bigg) =\textrm{constant}&, \\\nonumber\\
    \Omega_R^2(t) R_R^3(t) + \Omega_R(t)\big(5S_l + 3\Delta\Sigma_l \big) =\textrm{constant}=M,&\\\nonumber\\
    \Phi_R(t) + \frac{1}{32M^{5/2}\nu} R_R^{5/2}(t) -\frac{5\Big(41S_l + 15\Delta \Sigma_l \Big)}{256\nu M^{2}\mathscr{S}^{2}} \Bigg(\frac{64\nu M^3}{5}t +\frac{1}{4}R_R^4(t) + \frac{2\mathscr{S}}{5M^{1/2}}R_R^{5/2}(t)\Bigg) =\textrm{constant},&\\\nonumber\\
    A^S_R(t) -\frac{5A_R\Big(7S_l + 3\Delta \Sigma_l \Big)}{64\nu M^{2}\mathscr{S}^{2}}\Bigg(\frac{64\nu M^3}{5}t +\frac{1}{4}R_R^4(t) + \frac{2\mathscr{S}}{5M^{1/2}}R_R^{5/2}(t)  \Bigg) =\textrm{constant}&.
 \end{align} 
\end{subequations}

Note that one constraint appears in the RG equations of $R_R(t)$, (\ref{RGE_R}), which indicates the range of effectiveness of the DRG method, 
 \begin{align}
    \frac{\ud R_R}{\ud \tau} =& -\frac{64 \nu  M^3}{5 R_R^3} + \frac{64 \nu  M^{5/2} \mathscr{S}}{5 R_R^{9/2}} + \mathcal{O}(S^2) = -\frac{64 \nu  M^{5/2} }{5 R_R^{9/2}} \Big(M^{1/2}R_R^{3/2} -\mathscr{S}\Big).
 \end{align}
If $\mathscr{S} = (51 S_l + 21\Delta\Sigma_l)/4$ is positive, $R_R(t)$, which is the dominant part of the binary center-of-mass separation $r(t)$, decreases until $R_R(t) = \mathscr{S}^{2/3}M^{-1/3}$. Given a limitation on the smallest value of $R_R(t)$ and combining with (\ref{Renorm_Sol_R}), it is possible to determine an approximate end time of the inspiral phase described by the Post-Newtonian equations of motion (\ref{a_sum}). This could provide useful information to numerical simulations as well.\\

\section{Spin Precession Equations}
\label{appendix_prec_solutions}

In this section, we aim to obtain the analytic solutions for the spin precession equations at  linear order in spin  (\ref{spin_prec_decomp}) by applying DRG methods, with the quasi-circular solutions to the equations of motion from the previous section. For a conservative binary system moving in nearly circular motion, solving equations in the form of
\begin{align}
    \frac{\ud S^{a}_{n}}{\ud t} = (\Omega - \Omega_a  ) S^{a}_{\lambda}, \qquad \frac{\ud S^{a}_{\lambda}}{\ud t} = -(\Omega - \Omega_a  ) S^{a}_{n}, \qquad \textrm{with} \quad \Omega_{a} = \frac{\nu M\Omega}{R}\Big(2 + \frac{3}{2}\frac{m_b}{m_a}\Big),
\end{align}
is fairly straightforward for constant radius $R$ and orbital frequency $\Omega$. The solutions are $S^a_n = S^a_{\parallel} \sin{\Big((\Omega - \Omega_a)(t-t_0) + \Phi\Big)}$ and $S^a_\lambda = S^a_{\parallel} \cos{\Big((\Omega - \Omega_a)(t-t_0) + \Phi\Big)}$, where $S^a_{\parallel}$ is determined by the initial spin vectors.

With the inclusion of the radiation reaction force and the resulting time-dependence of $r(t)$ and $\omega(t)$, the spin vectors precess in a way entangled with the orbit motion. Defining $S^{a}_{+} \equiv S^{a}_{n} + i S^{a}_{\lambda}$,  the precession equations (\ref{spin_prec_decomp}) can be combined and written as
\begin{align}
    \frac{\ud S^{a}_{+}(t)}{\ud t} = -i \big(\omega (t) - \Omega_a(t) \big) S^{a}_{+}(t) .\label{Splus_evol}
\end{align}
A simple integration with respect to time leads to
\begin{equation}
    i\Big[\ln S^{a}_{+}(t) - \ln S^{a}_{+}(t_0) \Big] = \int^{t}_{t_0} \ud \tau \big[\omega(t) -\Omega_a(t) \big].\label{lnS_integrate}
\end{equation}
To solve for the integral on the right-hand side, we denote $ \nu_a\equiv (2+\frac{3m_b}{2m_a}) \nu^2  $ and recall that $M \sim \Omega_B^2 R_B^3 + \Omega_B(5S_l+3\Delta\Sigma_l)$, such that $\Omega_a$ in (\ref{Omega_a}) at leading order in spin becomes 
\begin{align}
    \Omega_a(t) &= \frac{\nu_a}{\nu} \Big(\Omega_B^2 R_B^3  + \Omega_B(5S_l+3\Delta\Sigma_l)\Big) \Bigg[\frac{\Omega_B}{R_B} + \frac{\delta \omega}{R_B} -\frac{\Omega_B \delta r}{R_B^2} \Bigg] + \frac{\nu_a}{\nu}\Omega_B^3 R_B^2\Bigg(\frac{\delta \omega_{S}}{\Omega_B} -\frac{\delta r_{S}}{R_B} \Bigg)  + \mathcal{O}(S^{2})\label{Omega_a_LS},
\end{align}
with the 2.5PN radiation perturbation $\{\delta r, \delta\omega\}$ from (\ref{delta_RR}), and the leading order spin-orbit perturbation $\{\delta r_S, \delta\omega_S\}$ from (\ref{del_pert_S}). 

As a check of self-consistency, notice that we have the choice of substituting $M$ either as a function of the physical values $\{r(t),\omega(t)\}$ using the results from (\ref{delta_r_omega}), or the bare parameters $\{R_B,\Omega_B\}$, which give the same result after summing up the perturbation expansions.

Substituting the corresponding perturbations back into (\ref{Omega_a_LS}), we obtain the explicit time-dependence of the precession norm $\Omega_a$, 
\begin{align}
    \Omega_a (t) = \frac{\nu_a}{\nu} &\Bigg[\Omega_B^3 R_B^2 + (5S_l+3\Delta\Sigma_l)\frac{\Omega_B^2}{R_B} + 32\nu R_B^7\Omega_B^9(t-t_0) +(184S_l + \frac{936}{5}\Delta\Sigma_l)\nu R_B^4\Omega_B^8(t-t_0)  \nonumber\\
    &-3 A_B\Omega_B^3  R_B \sin \left(\Omega_B(t-t_0) + \Phi_B \right) - \Big(\frac{13}{2}S_l +\frac{9}{2}\Delta\Sigma_l\Big)\frac{A_B\Omega_B^2 }{R_B^2} \sin \left(\Omega_B(t-t_0) + \Phi_B \right)  \nonumber\\
    & - (21S_l+9\Delta\Sigma_l)\frac{A_B \Omega_B^3}{R_B^2}(t-t_0)\cos \left(\Omega_B(t-t_0) + \Phi_B \right) -3 A_B^S \Omega_B^3  R_B \cos \left(\Omega_B(t-t_0) + \Phi_B \right)  \Bigg].
\end{align}
Combined with the expression for $\delta\omega(t)$ in terms of the time-independent bare parameters, we can perform the integration in (\ref{lnS_integrate}) to write down
\begin{align}
    i\Big[&\ln S^{a}_{+}(t) - \ln S^{a}_{+}(t_0) \Big] \nonumber\\
    =&\Bigg(\Omega_B - \frac{\nu_a}{\nu} \Omega_B^3 R_B^2  - \frac{\nu_a}{\nu}(5S_l+3\Delta\Sigma_l)\frac{\Omega_B^2}{R_B}\Bigg)(t - t_0) 
    \nonumber\\
    &+ \Bigg(\frac{48\nu}{5}R_B^5 \Omega_B^7 - \Big(\frac{12}{5}S_l - \frac{108}{5}\Delta \Sigma_l \Big)\nu R^2_B \Omega_B^6 - 16\nu_a R_B^7\Omega_B^9 -(92S_l + \frac{468}{5}\Delta\Sigma_l)\nu_a R_B^4\Omega_B^8\Bigg)(t -t_0)^2 
    \nonumber\\
    &+ \Bigg(\frac{2 A_B}{R_B} - \Big(5S_l + 3\Delta \Sigma_l \Big)\frac{A_B}{R_B^4\Omega_B}- \frac{3\nu_a}{\nu} \Omega_B^2 A_B R_B
    \nonumber\\
    &\phantom{++\Bigg(}- \frac{\nu_a}{\nu}\Big(\frac{13}{2}S_l +\frac{9}{2}\Delta\Sigma_l\Big)\frac{A_B\Omega_B }{R_B^2}\Bigg)\big[\cos{\left(\Omega_B(t-t_0) + \Phi_B \right)} - \cos\Phi_B\big] 
    \nonumber \\
    &-\Bigg(\Big(14S_l + 6\Delta \Sigma_l \Big)\frac{A_B}{R_B^4\Omega_B} - \frac{\nu_a}{\nu}(21S_l+9\Delta\Sigma_l)\frac{A_B \Omega_B}{R_B^2} \Bigg)   \nonumber \\
    &\phantom{--}\times
    \Big[\Omega_B(t-t_0)\sin \big(\Omega_B(t-t_0) + \Phi_B \big) + \cos{\left(\Omega_B(t-t_0) + \Phi_B \right)} - \cos\Phi_B \Big] 
    \nonumber\\
    &-\Bigg(\frac{2A^S_B }{R_B} - \frac{3\nu_a}{\nu} A_B^S \Omega_B^2  R_B\Bigg)\Big[\sin{(\Omega_B(t-t_0)+\Phi_B)} -\sin\Phi_B \Big].  
\end{align}
\normalsize
The constant terms $\sin\Phi_B$ and $\cos\Phi_B$ can be absorbed by redefining the initial condition $i\ln {S}^{a}_{+}(t_0)$, or via a bare parameter $i\ln \mathcal{S}^{a}_{+B}$,

\begin{align}
     i\ln {S}^{a}_{+}(t_0) \rightarrow i\ln \mathcal{S}^{a}_{+B} &+ \Bigg(\frac{2 A_B}{R_B} - \Big(5S_l + 3\Delta \Sigma_l \Big)\frac{A_B}{R_B^4\Omega_B}- \frac{3\nu_a}{\nu} \Omega_B^2 A_B R_B- \frac{\nu_a}{\nu}\Big(\frac{13}{2}S_l +\frac{9}{2}\Delta\Sigma_l\Big)\frac{A_B\Omega_B }{R_B^2}\Bigg)\cos \Phi_B \nonumber\\
     &-\Bigg(\Big(14S_l + 6\Delta \Sigma_l \Big)\frac{A_B}{R_B^4\Omega_B} - (21S_l+9\Delta\Sigma_l)\frac{\nu_a}{\nu}\frac{A_B \Omega_B}{R_B^2} \Bigg)\cos \Phi_B \nonumber\\
     &-\Bigg(\frac{2A^S_B }{R_B} - \frac{3\nu_a}{\nu} A_B^S \Omega_B^2  R_B\Bigg)\sin\Phi_B.
 \end{align} 
The logarithm of the spin components then becomes
\begin{align}
    i\ln S^{a}_{+}(t) &= i\ln \mathcal{S}^{a}_{+B} + \Bigg(\Omega_B - \frac{\nu_a}{\nu} \Omega_B^3 R_B^2  - \frac{\nu_a}{\nu}(5S_l+3\Delta\Sigma_l)\frac{\Omega_B^2}{R_B}\Bigg)(t - t_0) \nonumber\\
    &+ \Bigg(\frac{48\nu}{5}R_B^5 \Omega_B^7 - \Big(\frac{12}{5}S_l - \frac{108}{5}\Delta \Sigma_l \Big)\nu R^2_B \Omega_B^6 - 16\nu_a R_B^7\Omega_B^9 -(92S_l + \frac{468}{5}\Delta\Sigma_l)\nu_a R_B^4\Omega_B^8\Bigg)(t -t_0)^2 \nonumber\\
    &+ \Bigg(\frac{2 A_B}{R_B} - \Big(19S_l + 9\Delta \Sigma_l \Big)\frac{A_B}{R_B^4\Omega_B}- \frac{3\nu_a}{\nu} \Omega_B^2 A_B R_B+ \frac{\nu_a}{\nu}\Big(\frac{29}{2}S_l +\frac{9}{2}\Delta\Sigma_l\Big)\frac{A_B\Omega_B }{R_B^2}\Bigg)\cos{\left(\Omega_B(t-t_0) + \Phi_B \right)} \nonumber \\
    &-\Bigg(\Big(14S_l + 6\Delta \Sigma_l \Big)\frac{A_B}{R_B^4} - \frac{\nu_a}{\nu}(21S_l+9\Delta\Sigma_l)\frac{A_B \Omega_B^2}{R_B^2} \Bigg) (t-t_0)\sin \big(\Omega_B(t-t_0) + \Phi_B \big) \nonumber\\
    &-\Bigg(\frac{2A^S_B }{R_B} - \frac{3\nu_a}{\nu} A_B^S \Omega_B^2  R_B\Bigg)\sin{(\Omega_B(t-t_0)+\Phi_B)}.
    \label{lnS_bare}
\end{align}
\normalsize
Given the spin vector expansions in terms of the bare parameters $\{R_B, \Omega_B, \Phi_B, A_B, A^S_B, \mathcal{S}^{a}_{+B} \}$, the next step is to renormalize the spin components by replacing the bare parameters by the renormalized ones plus counter-terms, and splitting $t-t_0 =(t-\tau)+(\tau-t_0)$ with a choice of an arbitrary renormalization scale $\tau$.

\subsection{Spin Renormalization}

To begin, the bare parameter $\ln \mathcal{S}^{a}_{+B}$ is related to the renormalized value $\ln \mathcal{S}^{a}_{+R}$ through
\begin{equation}
    \ln \mathcal{S}^{a}_{+B}(t_0) = \ln \mathcal{S}^{a}_{+R}(\tau) + \delta_{\ln S}^a(\tau, t_0).
\end{equation}
The renormalization treatment is performed for the natural logarithm of the spin components. As a result, $\mathcal{S}^{a}_{+B} = \mathcal{S}^{a}_{+R} e^{\delta_{\ln S}^a}$. The exponential implies that it is the phase of the precession that is renormalized. Dividing the bare parameters into the renormalized parts and the counter-terms and introducing the renormalization scale $\tau$, Eq.~(\ref{lnS_bare}) then becomes
\begin{align}
    &i\Big[\ln S^{a}_{+}(t) - \Big(\ln \mathcal{S}^{a}_{+R} + \delta_{\ln S}^a \Big)\Big] \nonumber\\
    &=  \Bigg(\Omega_R + \delta_{\Omega} - \frac{\nu_a}{\nu} \Big(\Omega_R^3 R_R^2+3\Omega_R^2 R_R^2\delta_{\Omega} + 2\Omega_R^3 R_R\delta_R\Big)  - \frac{\nu_a}{\nu}(5S_l+3\Delta\Sigma_l)\Big(\frac{\Omega_R^2}{R_R} + \frac{2\Omega_R\delta_{\Omega}}{R_R}  -\frac{\Omega_R^2\delta_R}{R_R^2} \Big) \Bigg)
    \nonumber\\
    &\qquad\times\big[(t -\tau)+(\tau- t_0)\big] 
    \nonumber\\
    &+ \Bigg(\frac{48\nu}{5}R_R^5 \Omega_R^7 - \Big(\frac{12}{5}S_l - \frac{108}{5}\Delta \Sigma_l \Big)\nu R^2_R \Omega_R^6 - 16\nu_a R_R^7\Omega_R^9 -(92S_l + \frac{468}{5}\Delta\Sigma_l)\nu_a R_R^4\Omega_R^8\Bigg)
    \nonumber\\
    &\qquad\times\big[(t-\tau)^2 + 2(t-\tau)(\tau -t_0) +(\tau -t_0)^2 \big] 
    \nonumber\\
    &+ \Bigg(\frac{2 A_R}{R_R} - \Big(19S_l + 9\Delta \Sigma_l \Big)\frac{A_R}{R_R^4\Omega_R}- \frac{3\nu_a}{\nu} \Omega_R^2 A_R R_R+ \frac{\nu_a}{\nu}\Big(\frac{29}{2}S_l +\frac{9}{2}\Delta\Sigma_l\Big)\frac{A_R\Omega_R }{R_R^2}\Bigg)\cos{\left(\Omega_R(t-\tau) + \Phi_R \right)} 
    \nonumber \\
    &-\Bigg(\Big(14S_l + 6\Delta \Sigma_l \Big)\frac{A_R}{R_R^4} - \frac{\nu_a}{\nu}(21S_l+9\Delta\Sigma_l)\frac{A_R \Omega_R^2}{R_R^2} \Bigg) \big[(t-\tau)+(\tau-t_0)\big]\sin \big(\Omega_R(t-\tau) + \Phi_R \big) 
    \nonumber\\
    &-\Bigg(\frac{2A^S_R }{R_R} +\frac{2\delta_A^S }{R_R} - \frac{3\nu_a}{\nu} A_R^S \Omega_R^2  R_R- \frac{3\nu_a}{\nu} \delta_A^S \Omega_R^2  R_R\Bigg)\sin{(\Omega_R(t-\tau)+\Phi_R)}. \label{renorm_expan}
\end{align}
The counter-terms are the combined results in \cite{Galley:2016zee} and (\ref{counter_S}),
\begin{align*}
    \delta_{R}(\tau,t_0) &= \frac{64\nu}{5} R_R^6\Omega_R^6(\tau-t_0) +\delta^{S}_{R}(\tau,t_0),\\
    \delta_{\Omega}(\tau,t_0) &= - \frac{96\nu}{5} R_R^5\Omega_R^7(\tau-t_0)  +\delta^{S}_\Omega(\tau,t_0),\\
    \delta_{\Phi}(\tau,t_0) &= -\Omega_R(\tau-t_0)+\frac{48\nu}{5} R_R^5\Omega_R^7(\tau-t_0)^2 +\delta^{S}_\Phi(\tau,t_0). 
\end{align*}
After some algebra, (\ref{renorm_expan}) can be simplified to
\begin{align}
    &i\Big[\ln S^{a}_{+}(t) - \Big(\ln \mathcal{S}^{a}_{+R} + \delta_{\ln S}^a \Big)\Big] \nonumber\\
    = &\Big(\Omega_R -\frac{\nu_a}{\nu}\Omega_R^3 R_R^2 - \frac{\nu_a}{\nu}(5S_l+3\Delta\Sigma_l)\frac{\Omega_R^2}{R_R}\Big) (t-\tau) + \Big(\Omega_R -\frac{\nu_a}{\nu}\Omega_R^3 R_R^2  - \frac{\nu_a}{\nu}(5S_l+3\Delta\Sigma_l)\frac{\Omega_R^2}{R_R}\Big)(\tau-t_0)\nonumber\\
    &+ \Bigg(\frac{48\nu}{5}R_R^5 \Omega_R^7 - \Big(\frac{12}{5}S_l - \frac{108}{5}\Delta \Sigma_l \Big)\nu R^2_R \Omega_R^6 - 16\nu_a R_R^7\Omega_R^9 -(92S_l + \frac{468}{5}\Delta\Sigma_l)\nu_a R_R^4\Omega_R^8\Bigg)\big[(t-\tau)^2 -(\tau -t_0)^2 \big] \nonumber\\
    &+ \Bigg(\frac{2 A_R}{R_R} - \Big(19S_l + 9\Delta \Sigma_l \Big)\frac{A_R}{R_R^4\Omega_R}- \frac{3\nu_a}{\nu} \Omega_R^2 A_R R_R+ \frac{\nu_a}{\nu}\Big(\frac{29}{2}S_l +\frac{9}{2}\Delta\Sigma_l\Big)\frac{A_R\Omega_R }{R_R^2}\Bigg)\cos{\left(\Omega_R(t-\tau) + \Phi_R \right)} \nonumber \\
    &-\Bigg(\Big(14S_l + 6\Delta \Sigma_l \Big)\frac{A_R}{R_R^4} - \frac{\nu_a}{\nu}(21S_l+9\Delta\Sigma_l)\frac{A_R \Omega_R^2}{R_R^2} \Bigg) (t-\tau)\sin \big(\Omega_R(t-\tau) + \Phi_R \big) \nonumber\\
    &-\Bigg(\frac{2A^S_R }{R_R} - \frac{3\nu_a}{\nu} A_R^S \Omega_R^2  R_R\Bigg)\sin{(\Omega_R(t-\tau)+\Phi_R)}. 
\end{align}
Notice that the terms proportional to $(t-\tau)(\tau-t_0)$  are completely canceled, which was emphasized in \cite{Galley:2016zee} as an important check of self-consistency. Here the cancellation is due to exactly the same set of substitutions we could use to replace $M$ to obtain (\ref{Omega_a_LS}), where the two different choices led to the same expansion result. 

To cancel the remaining secular pieces that are proportional to the powers of $(\tau-t_0)$, the counter-term $\delta_{\ln S}^a$ is fixed to be 
\begin{align}
    i\delta_{\ln S}^a(\tau,t_0) &= -\Big(\Omega_R -\frac{\nu_a}{\nu}\Omega_R^3 R_R^2  - \frac{\nu_a}{\nu}(5S_l+3\Delta\Sigma_l)\frac{\Omega_R^2}{R_R}\Big)(\tau-t_0) \nonumber\\
    &+ \Bigg(\frac{48\nu}{5}R_R^5 \Omega_R^7 - \Big(\frac{12}{5}S_l - \frac{108}{5}\Delta \Sigma_l \Big)\nu R^2_R \Omega_R^6 - 16\nu_a R_R^7\Omega_R^9 -(92S_l + \frac{468}{5}\Delta\Sigma_l)\nu_a R_R^4\Omega_R^8\Bigg)(\tau-t_0)^2. \label{delta_lnS}
\end{align}
Choosing the arbitrary scale $\tau$ to equal $t$, the renormalized solution to $\ln \mathcal{S}^{a}_{+}(t)$ becomes
\begin{align}
    \label{spin_resummed}
    i\ln S^{a}_{+}(t) =& i\ln \mathcal{S}^{a}_{+R}-\Bigg(\frac{2A^S_R }{R_R} - \frac{3\nu_a}{\nu} A_R^S \Omega_R^2  R_R\Bigg)\sin{\Phi_R} \nonumber\\
    +& \Bigg(\frac{2 A_R}{R_R} - \Big(19S_l + 9\Delta \Sigma_l \Big)\frac{A_R}{R_R^4\Omega_R}- \frac{3\nu_a}{\nu} \Omega_R^2 A_R R_R+ \frac{\nu_a}{\nu}\Big(\frac{29}{2}S_l +\frac{9}{2}\Delta\Sigma_l\Big)\frac{A_R\Omega_R }{R_R^2}\Bigg)\cos{\Phi_R }.
\end{align}
or more explicitly in terms of the exponential,
\begin{align}
    S^{a}_{+}(t) = \mathcal{S}^{a}_{+R} &\exp \Bigg\{i\Bigg(\frac{2A^S_R }{R_R} - \frac{3\nu_a}{\nu} A_R^S \Omega_R^2  R_R\Bigg)\sin{\Phi_R}\nonumber\\
    &-i\Bigg(\frac{2 A_R}{R_R} - \Big(19S_l + 9\Delta \Sigma_l \Big)\frac{A_R}{R_R^4\Omega_R}- \frac{3\nu_a}{\nu} \Omega_R^2 A_R R_R+ \frac{\nu_a}{\nu}\Big(\frac{29}{2}S_l +\frac{9}{2}\Delta\Sigma_l\Big)\frac{A_R\Omega_R }{R_R^2}\Bigg)\cos{\Phi_R }\Bigg\} .
\end{align}
The renormalized quantities as functions of time have runnings obtained from the RG flow in Section \ref{EoM_RGE}, with only the remaining the spin component bare parameter $\mathcal{S}^{a}_{+R}$ to be done in the next section.

\subsection{Spin Component Renormalization Group Solution}

The running of the renormalized parameter $\mathcal{S}^{a}_{+R}$ can be determined using (\ref{delta_lnS}), which leads to
\begin{align}
    \frac{\ud }{\ud \tau} i\ln \mathcal{S}^{a}_{+R}(\tau)&=  \Big(\Omega_R -\frac{\nu_a}{\nu}\Omega_R^3 R_R^2  - \frac{\nu_a}{\nu}(5S_l+3\Delta\Sigma_l)\frac{\Omega_R^2}{R_R}\Big) \nonumber\\
    &+ \Bigg[\frac{\ud \Omega_R}{\ud \tau} - \frac{\nu_a}{\nu}\Omega_R^3 R_R^2 \Bigg(\frac{3}{\Omega_R}\frac{\ud \Omega_R}{\ud \tau}+ \frac{2}{R_R}\frac{\ud R_R}{\ud \tau}\Bigg) - \frac{\nu_a}{\nu}(5S_l+3\Delta\Sigma_l)\frac{\Omega_R^2}{R_R} \Bigg(\frac{2}{\Omega_R}\frac{\ud \Omega_R}{\ud \tau} - \frac{1}{R_R}\frac{\ud R_R}{\ud \tau}\Bigg)  \Bigg]
    \nonumber\\
    &\qquad\times(\tau -t_0)\nonumber\\
    &+ \Bigg[\frac{96\nu}{5}R_B^5 \Omega_B^7 - \Big(\frac{24}{5}S_l - \frac{216}{5}\Delta \Sigma_l \Big)\nu R^2_B \Omega_B^6 - 32\nu_a R_B^7\Omega_B^9 -(184S_l + \frac{936}{5}\Delta\Sigma_l)\nu_a R_B^4\Omega_B^8\Bigg]
    \nonumber\\
    &\qquad\times(\tau -t_0). \nonumber\\
\end{align}
\normalsize
It seems to be formally divergent and has the dependence on the cut-off $t_0$. However, replacing the derivatives of $R_R$ and $\Omega_R$ by their RG equations (\ref{RGE_R}) and (\ref{RGE_Omega}), we encounter the non-trivial cancellation and obtain a finite $\beta$-function,
\begin{align}
    \frac{\ud }{\ud \tau} i\ln \mathcal{S}^{a}_{+R}(\tau) &= \Omega_R -\frac{\nu_a}{\nu}\frac{M\Omega_R}{R_R}. \label{RGE_ln_S}
\end{align}
Notice the similarity in form between the RG equation and (\ref{Splus_evol}), the precession equation we start with. 

In order to find a solution to the RG equation of the spin component, we can write the relation between the $\tau$-derivative of $i\ln \mathcal{S}^{a}_{+R}$ and the derivative with respect to the renormalized parameter $R_R$ as 
\begin{align}
     \frac{\ud }{\ud R_R} i\ln \mathcal{S}^{a}_{+R}(\tau) =& \Bigg(\frac{\ud R_R}{\ud \tau}   \Bigg)^{-1}\frac{\ud }{\ud \tau} i\ln \mathcal{S}^{a}_{+R}(\tau).
\end{align} 
Using the RGEs (\ref{RGE_R}) and (\ref{RGE_ln_S}), we obtain a solution to $\mathcal{S}^{a}_{+R}(\tau)$ in terms of $R_R(\tau)$ and initial conditions
\begin{align}
\label{Renorm_Sol_S}
    i\ln \mathcal{S}^{a}_{+R}(t) =& i\ln \mathcal{S}^{a}_{+R}(t_i) + \Big(\Phi_R(t) - \Phi_R(t_i)\Big) + \frac{5\nu_a}{96M^{3/2}\nu^2}\Big[R^{3/2}_R(t)-R^{3/2}_R(t_i)  \Big] \nonumber\\
    &+ \frac{5\Big(41S_l + 15\Delta \Sigma_l \Big)\nu_a}{384M^2\nu^2}\Big[\ln\big(M^{1/2}R_R^{3/2}(t) - \mathscr{S}\big) -\ln\big(M^{1/2}R_R^{3/2}(t_i) - \mathscr{S}\big) \Big].
\end{align}
The expressions are not unique in terms of $\Phi_R(t)$ and $R_R(t)$ due to several RG invariants between them. The invariance over time with spin components can be found from the $\mathcal{S}^{a}_{+R}(\tau)$ solution, which is given by
\begin{align}
    i\ln \mathcal{S}^{a}_{+R}(t) - \Phi_R(t) -\frac{5\nu_a R^{3/2}_R(t)}{96M^{3/2}\nu^2} - \frac{5\Big(41S_l + 15\Delta \Sigma_l \Big)\nu_a}{384M^2\nu^2}\ln\big(M^{1/2}R_R^{3/2}(t) - \mathscr{S}\big)=\textrm{constant}.
\end{align}
Putting all the pieces together, the resummed solution of $S^{a}_{+}(t)$ is given by
\begin{align}
    S^{a}_{+}(t) = \mathcal{S}^{a}_{+R}(t_i) \times \Bigg(&\frac{M^{1/2}R_R^{3/2}(t_i) - \mathscr{S}}{M^{1/2}R_R^{3/2}(t) - \mathscr{S}}\Bigg)^{\frac{5i(41S_l + 15\Delta \Sigma_l)\nu_a}{(384M^2\nu^2)}} \times \exp \Bigg\{-i\Bigg[\Big(\Phi_R(t) - \Phi_R(t_i)\Big) + \frac{5\nu_a\Big(R^{3/2}_R(t)-R^{3/2}_R(t_i)  \Big)}{96M^{3/2}\nu^2} \Bigg] \nonumber \\
    &+ i\Bigg(\frac{2A^S_R (t)}{R_R(t)} - \frac{3\nu_a}{\nu} A_R^S(t) \Omega_R(t)^2  R_R(t)\Bigg)\sin{\Phi_R(t)}-i\Bigg[\frac{2 A_R(t)}{R_R(t)} - \Big(19S_l + 9\Delta \Sigma_l \Big)\frac{A_R(t)}{R_R(t)^4\Omega_R(t)}\nonumber\\
    &\qquad\qquad- \frac{3\nu_a}{\nu} \Omega_R(t)^2 A_R(t) R_R(t)+ \frac{\nu_a}{\nu}\Big(\frac{29}{2}S_l +\frac{9}{2}\Delta\Sigma_l\Big)\frac{A_R(t)\Omega_R(t) }{R_R(t)^2}\Bigg]\cos{\Phi_R(t) }\Bigg\},\label{S_a_t}
\end{align}
with $\{R_R(t), \Omega_R(t), \Phi_R(t), A_R(t), A^S_R(t), \mathcal{S}^{a}_{+R}(t) \}$ given by (\ref{Renorm_Sol}), (\ref{Renorm_Sol_R}) and (\ref{Renorm_Sol_S}).

The quantity $\mathcal{S}^{a}_{+R}(t_i)$ depends on the initial conditions of dynamics and spin vectors. For instance, taking the initial input $S^a_{n}(t_i)$ and $S^a_{\lambda}(t_i)$, while getting $A_R(t_i)$, $R_R(t_i)$, $\Omega_R(t_i)$ and $\Phi_R(t_i)$ from numerically solving the initial conditions $r(t_i)$, $\dot{r}(t_i)$, $\omega(t_i)$ and $\phi(t_i)$ from the dynamics, we can determine the value of $\mathcal{S}^{a}_{+R}(t_i)$ through

\begin{align}
     \mathcal{S}^{a}_{+R}(t_i) = \big(&S^a_{n}(t_i) + iS^a_{\lambda}(t_i) \big) \exp \Bigg\{i\Bigg[\frac{2 A_R(t_i)}{R_R(t_i)} - \Big(19S_l + 9\Delta \Sigma_l \Big)\frac{A_R(t_i)}{R_R^4(t_i)\Omega_R(t_i)} - \frac{3\nu_a}{\nu} \Omega_R^2(t_i) A_R(t_i) R_R(t_i) \nonumber\\
     &
     \phantom{\big(S^a_{n}(t_i) + iS^a_{\lambda}(t_i) \big) \exp \Bigg\{i\Bigg[+}
     + \frac{\nu_a}{\nu}\Big(\frac{29}{2}S_l +\frac{9}{2}\Delta\Sigma_l\Big)\frac{A_R(t_i)\Omega_R(t_i) }{R_R^2(t_i)}\Bigg)\cos{\Phi_R(t_i) } \Bigg] 
    \nonumber\\
    &
     \phantom{\big(S^a_{n}(t_i) + iS^a_{\lambda}(t_i) \big) \exp \Bigg\{}
     - i\Bigg(\frac{2A^S_R (t_i)}{R_R(t_i)} - \frac{3\nu_a}{\nu} A_R^S(t_i) \Omega_R^2(t_i)  R_R(t_i)\Bigg)\sin{\Phi_R(t_i)}\Bigg\}. \label{S_a_R_ti}
 \end{align} 
\normalsize

 One immediate validation of the formulation is that the length of the spin vector should be a constant. Thus $\left|S^{a}_{+}(t)\right| = \sqrt{(S^{a}_{n})^2 + (S^{a}_{\lambda})^2}$ should be a constant, since $S^{a}_{l}$ does not change with time. From (\ref{S_a_t}) and (\ref{S_a_R_ti}) we can see that the length is preserved, $\left|S^{a}_{+}(t)\right|=\left|S^{a}_{+R}(t_i)\right| = \sqrt{(S^{a}_{n}(t_i))^2 + (S^{a}_{\lambda}(t_i))^2}$ as long as $\big(M^{1/2}R_R^{3/2}(t) - \mathscr{S}\big)>0 $, the same constraint we encounter for the solutions of the orbit equations of motion.

\section{The Moving Triad Evolution}
\label{appendix_frame}

In the text, the resummed analytic expressions for the orbital equations of motion and spin precession we obtained are written in terms of the moving triad vectors $\{\bm{n}, \bm{\lambda}, \bm{l}\}$. To transform the complete results into a fixed frame, we follow the solutions to the evolution equations for the moving triad in \cite{Blanchet:2011zv, Blanchet:2013haa} for the 1.5PN order conservative dynamics and build the moving triad evolution for the radiative dynamics on quasi-circular orbits.

\begin{wrapfigure}{r}{0.4\textwidth}
  \vspace{-30pt}
  \begin{center}
    \includegraphics[trim=3cm 17.5cm 8cm 2.cm, clip,width=0.4\textwidth]{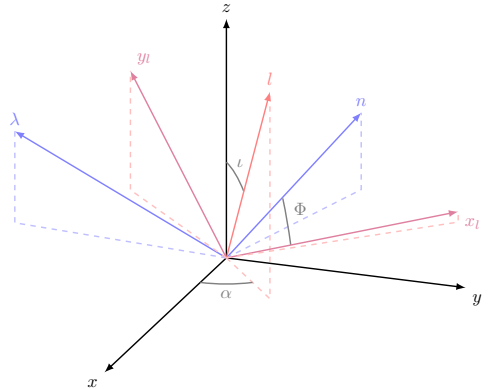}
  \end{center}
  \vspace{-20pt}
    \caption{Definitions of the Euler angle $\{\alpha, \iota, \Phi\}$ with respect to the moving triad $\{\bm{n}, \bm{\lambda}, \bm{l}\}$, the auxiliary moving frame $\{\bm{x}_l, \bm{y}_l, \bm{l}\}$, and the fixed lab frame $\{\bm{x},\bm{y},\bm{z}\}$.}
    \label{Euler_angle}
\end{wrapfigure}

We start by briefly summarizing the conservative moving triad evolution solution that relies fundamentally on the conservation of the total angular momentum $\bm{J}$. An orthonormal inertial frame $\{\bm{x},\bm{y},\bm{z}\}$ is then introduced with $\bm{J}/J$ as the fixed direction $\bm{z}$. Three Euler angles $\alpha(t), \iota(t), \Phi(t)$ are defined to specify the moving triad within the fixed frame as shown in Figure \ref{Euler_angle}. The azimuth $\alpha$ and the inclination $\iota$ are the standard spherical coordinates of the Newtonian angular momentum direction $\bm{l}$. The angle $\Phi$ is defined to be the angle between $\bm{n}$ and $\bm{x}_l$, where
\begin{align}
     \bm{x}_l = \frac{\bm{z}\times\bm{l}}{|\bm{z}\times\bm{l}|}, \qquad \bm{y}_l = \bm{l} \times \bm{x}_l,
 \end{align} 
forming the instantaneous orbital plane and with $\bm{l}$ to complete an auxiliary orthonormal basis \{$\bm{x}_{l} , \bm{y}_{l} , \bm{l}$\}.

In terms of the Euler angles, the relation between the moving triad $\{\bm{n}(t),\bm{\lambda}(t),\bm{l}(t)\}$ and the fixed Cartesian frame  $\{\bm{x},\bm{y},\bm{z}\}$ can be written as

\begin{subequations}
\label{n_lambda_l_in_xyz}
\begin{align}
    \bm{n}&= (-\cos\Phi\sin\alpha - \sin\Phi\cos\iota\cos\alpha ) \bm{x} + (\cos\Phi\cos\alpha - \sin\Phi\cos\iota\sin\alpha )\bm{y} +\sin\Phi\sin\iota \bm{z}, \\
    \bm{\lambda}&= (\sin\Phi\sin\alpha - \cos\Phi\cos\iota\cos\alpha ) \bm{x} + (-\sin\Phi\cos\alpha - \cos\Phi\cos\iota\sin\alpha )\bm{y} +\cos\Phi\sin\iota \bm{z}, \\
    \bm{l} &= \sin\iota\cos\alpha \bm{x} + \sin\iota\sin\alpha \bm{y} + \cos\iota \bm{z}.
\end{align}
\end{subequations}
The evolution solutions to the Euler angles up to  linear order in spin are given by the components of the total angular momentum $\bm{J} = J_n(t) \hat{\bm{n}} + J_{\lambda}(t) \hat{\bm{\lambda}} + J_l(t) \hat{\bm{l}}$ as
\begin{align}
    \Phi + \alpha = \phi,\qquad \sin\iota = \frac{\sqrt{J_{n}^2 + J_{\lambda}^2}}{J}, \qquad e^{i\alpha} = \frac{J_{\lambda} - i J_{n}}{J}e^{i\phi},
\end{align}
where $\phi$ is the orbital phase, for which the resummed solution is given by (\ref{phi_resummed}) for the radiative binary orbits. 

Finally, expressed in terms of some initial basis $\{\bm{n}_{0},\bm{\lambda}_{0},\bm{l}_{0}\}$ with corresponding Euler angles $\{\alpha_0,\iota_0,\Phi_0\}$, the moving triad $\{\bm{n}(t),\bm{\lambda}(t),\bm{l}(t)\}$ is given by
 \begin{align}
 \bm{m} = e^{-i(\phi - \phi_0)}\bm{m}_0 + \frac{i}{\sqrt{2}}\left(\sin\iota e^{i\alpha} - \sin\iota_0 e^{i\alpha_0} \right)e^{-i\phi}\bm{l}_0 + \mathcal{O}(S^2) \\
 \bm{l} =\bm{l}_0 + \left[\frac{i}{\sqrt{2}}\left(\sin\iota e^{-i\alpha} - \sin\iota_0 e^{-i\alpha_0} \right)e^{i\phi_0}\bm{m}_0 + \textrm{c.c}  \right]+ \mathcal{O}(S^2), \label{l_Sol}
 \end{align}
where $\bm{m} \equiv \frac{1}{\sqrt{2}}(\bm{n} + i\bm{\lambda})$ is a complex null vector.

The crucial point of this moving triad solution is the conservation of the total angular momentum and the ability to write out its components in the moving triad for all time, not the physical meaning to $\bm{J}$. In order to apply the triad solutions to a radiative motion where $\bm{J}$ can change, we find such a quantity that satisfies the requirements by observing the calculation of $\ud \bm{J} / \ud t$ for a conservative quasi-circular orbit. It is conventional to decompose $\bm{J} = \bm{L} + \bm{S}$, where $\bm{S}$ is the total spin specified by the choices of spin variables following \cite{Kidder:1995zr}, and $\bm{L}$ is the sum of the non-spinning Newtonian $\bm{L}_{\textrm{N}}$ and the leading order spin-orbit contribution $\bm{L}_{\textrm{SO}}$, given by
\begin{align}
    \bm{L}_{\textrm{SO}} = \nu \left\{\frac{M}{r}\bm{n}\times\left[\bm{n}\times\left(3\bm{S} + \frac{\delta m}{m}\bm{\Delta} \right) \right] - \frac{1}{2} \bm{v}\times\left[\bm{v}\times\left(\bm{S} + \frac{\delta m}{m}\bm{\Delta} \right) \right] \right\}.
\end{align}
Written in terms of the moving triad components and taking the orbit radius and frequency as constants $R$ and $\Omega$ for the quasi-circular approximation, the spin-orbit momentum becomes
\begin{align}
    \bm{L}_{\textrm{SO}} =&  \frac{1}{2}\nu R^2\Omega^2 \left(\frac{m_2}{m_1} S^1_{n} + \frac{m_1}{m_2} S^2_{n} \right) \bm{n} -\frac{\nu M}{R}\left(\Big(\frac{m_2}{m_1}+2\Big) S^1_{\lambda} + \Big(\frac{m_1}{m_2}+2\Big) S^2_{\lambda} \right) \bm{\lambda} \nonumber\\
    &+  \left[ \frac{1}{2}\nu R^2\Omega^2 \left(\frac{m_2}{m_1} S^1_{l} + \frac{m_1}{m_2} S^2_{l} \right) -\frac{\nu M}{R}\left(\Big(\frac{m_2}{m_1}+2\Big) S^1_{l} + \Big(\frac{m_1}{m_2}+2\Big) S^2_{l} \right)\right]\bm{l}.
\end{align}
For a conservative system without radiation, the time derivative to the sum $\bm{J} = \bm{L}_\textrm{N} + \bm{L}_\textrm{SO} + \bm{S}_1 + \bm{S}_2 $ should vanish up to the Newtonian and leading spin order. By carrying out the detail calculation, we find that 
\begin{align*}
    \dot{\bm{L}}_{\textrm{SO}} = \Bigg[&\frac{1}{2}\nu R^2\Omega^2 \left(\frac{m_2}{m_1} \dot{S}^1_{n} + \frac{m_1}{m_2} \dot{S}^2_{n} \right) + \frac{\nu M\Omega}{R}\left(\Big(\frac{m_2}{m_1}+2\Big) S^1_{\lambda} + \Big(\frac{m_1}{m_2}+2\Big) S^2_{\lambda} \right) \Bigg]\bm{n} \nonumber\\
     &+ \Bigg[\frac{1}{2}\nu R^2\Omega^3 \left(\frac{m_2}{m_1} S^1_{n} + \frac{m_1}{m_2} S^2_{n} \right)  -\frac{\nu M}{R}\left(\Big(\frac{m_2}{m_1}+2\Big) \dot{S}^1_{\lambda} + \Big(\frac{m_1}{m_2}+2\Big) \dot{S}^2_{\lambda}\right)\Bigg]\bm{\lambda},\\
    \dot{\bm{S}} = -\frac{\nu M \Omega}{R}&\Big[\Big(2+\frac{3m_2}{2m_1} \Big) S^1_{\lambda} + \Big(2+\frac{3m_1}{2m_2} \Big) S^2_{\lambda} \Big]\bm{n} + \frac{\nu M \Omega}{R} \Big[\Big(2+\frac{3m_2}{2m_1} \Big) S^1_{n} + \Big(2+\frac{3m_1}{2m_2} \Big) S^2_{n} \Big]\bm{\lambda}, \\
    \dot{\bm{L}}_\textrm{N} = -\frac{\nu M \Omega}{R}\Big[&\Big(4+\frac{3m_2}{m_1} \Big) S^1_{n} + \Big(4+\frac{3m_1}{m_2} \Big) S^2_{n} \Big]\bm{\lambda}.
\end{align*}
\normalsize
Thus the sum is
\begin{align}
    \dot{\bm{L}}_\textrm{N} +\dot{\bm{L}}_{\textrm{SO}}  + \dot{\bm{S}} = -\frac{\nu M}{R}\Bigg[\Bigg(\frac{1}{2}\frac{m_1}{m_2}\Omega_2 S^2_{\lambda} + \frac{1}{2}\frac{m_2}{m_1}\Omega_1S^1_{\lambda}\Bigg) \bm{n} + \Bigg(\Big(2+\frac{m_1}{m_2}\Big)\Omega_2 S^2_{n} + \Big(2+\frac{m_2}{m_1}\Big)\Omega_1 S^1_{n}\Bigg)\bm{\lambda}\Bigg] \sim \mathcal{O}(v^2) \label{v^5S},
\end{align}
which correspond to 1PN terms to be fixed by including higher-order orbital angular momenta. Notice that the time derivative of the spins in $\dot{\bm{L}}_{\textrm{SO}}$ are completely canceled by $\dot{\bm{S}}$ and $\dot{\bm{L}}_\textrm{N}$ at  Newtonian order. Therefore we propose that for a radiative quasi-circular binary, the following quantity is conserved:
\begin{align}
    \mathscr{J} =&  \sum_{a,b}\Bigg\{\frac{1}{2}\nu \frac{m_b}{m_a} r(0)^2\omega(0)^2  S^a_{n}(t) \bm{n} -\frac{\nu M}{r(0)}\Big(\frac{m_b}{m_a}+2\Big) S^a_{\lambda}(t) \bm{\lambda} \nonumber +  \left[ \frac{1}{2}\nu \frac{m_b}{m_a} r(0)^2\omega(0)^2  S^a_{l}(t)  -\frac{\nu M}{r(0)}\Big(\frac{m_b}{m_a}+2\Big) S^a_{l}(t) \right]\bm{l} \Bigg\} \nonumber \\
    &+\sum_{a,b} \nu M r(0)^2 \omega(0) \bm{l} + \sum_{a,b} \bm{S}^a(t).
\end{align}
Compared with the conservative expressions, we replace the constant orbital radius and frequency by the initial orbital radius and frequency. The conservative spin components are changed into the time-dependent resummed radiative spin component results. The time derivative of this quantity $\mathscr{J}$ is $\sim \mathcal{O}(v^4S)$ but we are able to avoid the loss of total angular momentum due to non-spinning radiation at $\mathcal{O}(v^5)$. Using the substitution with $\mathscr{J}$ instead of $\bm{J}$ into the moving frame solutions (\ref{n_lambda_l_in_xyz}-\ref{l_Sol}), we can generate 3D-plot of the orbital radius evolution and animations of binary inspiral with spin orientation at every instant.

\bibliography{LSO}{}

\end{document}